\documentclass[twocolumn]{aastex63}

\usepackage{breqn}
\usepackage{amsmath}
\usepackage{rotating}
\usepackage{tikz}

\newcommand\kms{\ifmmode{\rm km\thinspace s^{-1}}\else km\thinspace s$^{-1}$\thinspace \fi}
\usepackage[version=4]{mhchem}

\received{XX}
\revised{YY}
\accepted{ZZ}
\submitjournal{AJ}
\shorttitle{MAPS XVI}
\shortauthors{Booth et al.}
\graphicspath{{./}{figures/}}
\usepackage{threeparttable}

\begin{document}

\title{Molecules with ALMA at Planet-forming Scales (MAPS) XVI: Characterizing the impact of the molecular wind on the evolution of the HD~163296 system}

\correspondingauthor{Alice S. Booth}
\email{abooth@strw.leidenuniv.nl}

\received{February 22, 2021}
\revised{June 16, 2021}
\accepted{July 30, 2021}

\author[0000-0003-2014-2121]{Alice S. Booth}
\affiliation{Leiden Observatory, Leiden University, 2300 RA Leiden, the Netherlands}
\affiliation{School of Physics and Astronomy, University of Leeds, Leeds, UK, LS2 9JT}
\author[0000-0002-1103-3225]{Beno\^{i}t Tabone}
\affil{Leiden Observatory, Leiden University, 2300 RA Leiden, the Netherlands}

\author[0000-0003-1008-1142]{John~D.~Ilee} 
\affiliation{School of Physics and Astronomy, University of Leeds, Leeds, UK, LS2 9JT}

\author[0000-0001-6078-786X]{Catherine~Walsh}
\affiliation{School of Physics and Astronomy, University of Leeds, Leeds, UK, LS2 9JT}

\author[0000-0003-3283-6884]{Yuri Aikawa}
\affiliation{Department of Astronomy, Graduate School of Science, The University of Tokyo, Tokyo 113-0033, Japan}

\author[0000-0003-2253-2270]{Sean M. Andrews}
\affiliation{Center for Astrophysics \textbar\, Harvard \& Smithsonian, 60 Garden Street, Cambridge, MA 02138, USA}

\author[0000-0001-7258-770X]{Jaehan Bae}
\altaffiliation{NASA Hubble Fellowship Program Sagan Fellow}
\affiliation{Earth and Planets Laboratory, Carnegie Institution for Science, 5241 Broad Branch Rd NW, Washington, DC 20015, USA}
\affiliation{Department of Astronomy, University of Florida, Gainesville, FL 32611, USA}

\author[0000-0003-4179-6394]{Edwin A.\ Bergin}
\affiliation{Department of Astronomy, University of Michigan, 323 West Hall, 1085 South University Avenue, Ann Arbor, MI 48109, USA}

\author[0000-0002-8716-0482]{Jennifer B. Bergner} \affiliation{University of Chicago, Department of the Geophysical Sciences, Chicago, IL 60637, USA}
\altaffiliation{NASA Hubble Fellowship Program Sagan Fellow}
 
\author[0000-0003-4001-3589]{Arthur D. Bosman}
\affiliation{Department of Astronomy, University of Michigan, 323 West Hall, 1085 South University Avenue, Ann Arbor, MI 48109, USA}

\author[0000-0002-0150-0125]{Jenny K. Calahan} 
\affiliation{Department of Astronomy, University of Michigan, 323 West Hall, 1085 South University Avenue, Ann Arbor, MI 48109, USA}

\author[0000-0002-2700-9676]{Gianni Cataldi}
\affiliation{National Astronomical Observatory of Japan, 2-21-1 Osawa, Mitaka, Tokyo 181-8588, Japan}
\affiliation{Department of Astronomy, Graduate School of Science, The University of Tokyo, Tokyo 113-0033, Japan}

\author[0000-0003-2076-8001]{L. Ilsedore Cleeves}
\affiliation{Department of Astronomy, University of Virginia, Charlottesville, VA 22904, USA}

\author[0000-0002-1483-8811]{Ian Czekala}
\affiliation{Department of Astronomy \& Astrophysics, 525 Davey Laboratory, The Pennsylvania State University, University Park, PA 16802, USA}
\affiliation{Center for Exoplanets \& Habitable Worlds, 525 Davey Laboratory, The Pennsylvania State University, University Park, PA 16802, USA}
\affiliation{Center for Astrostatistics, 525 Davey Laboratory, The Pennsylvania State University, University Park, PA 16802, USA}
\affiliation{Institute for Computational \& Data Sciences, The Pennsylvania State University, University Park, PA 16802, USA}
\affiliation{Department of Astronomy, 501 Campbell Hall, University of California, Berkeley, CA 94720-3411, USA}
\altaffiliation{NASA Hubble Fellowship Program Sagan Fellow}

\author[0000-0003-4784-3040]{Viviana V. Guzm\'{a}n}
\affiliation{Instituto de Astrof\'isica, Pontif\'ificia Universidad Cat\'olica de Chile, Av. Vicu\~na Mackenna 4860, 7820436 Macul, Santiago, Chile}

\author[0000-0001-6947-6072]{Jane Huang}
\altaffiliation{NASA Hubble Fellowship Program Sagan Fellow}
\affiliation{Department of Astronomy, University of Michigan, 323 West Hall, 1085 South University Avenue, Ann Arbor, MI 48109, USA}
\affiliation{Center for Astrophysics \textbar\, Harvard \& Smithsonian, 60 Garden Street, Cambridge, MA 02138, USA}

\author[0000-0003-1413-1776]{Charles J. Law}
\affiliation{Center for Astrophysics \textbar\, Harvard \& Smithsonian, 60 Garden Street, Cambridge, MA 02138, USA}

\author[0000-0003-1837-3772]{Romane Le Gal}
\affiliation{Center for Astrophysics \textbar\, Harvard \& Smithsonian, 60 Garden Street, Cambridge, MA 02138, USA}
\affiliation{IRAP, Universit\'{e} de Toulouse, CNRS, CNES, UT3, 31400 Toulouse, France}
\affiliation{IPAG, Universit\'{e} Grenoble Alpes, CNRS, IPAG, 38000 Grenoble, France}
\affiliation{IRAM, 300 rue de la piscine, F-38406 Saint-Martin d'H\`{e}res, France}

\author[0000-0002-7607-719X]{Feng Long}
\affiliation{Center for Astrophysics \textbar\, Harvard \& Smithsonian, 60 Garden Street, Cambridge, MA 02138, USA}

\author[0000-0002-8932-1219]{Ryan A. Loomis}
\affiliation{National Radio Astronomy Observatory, 520 Edgemont Road, Charlottesville, VA 22903, USA}

\author[0000-0002-1637-7393]{Fran\c cois M\'enard}
\affiliation{IRAP, Universit\'{e} de Toulouse, CNRS, CNES, UT3, 31400 Toulouse, France}
\author[0000-0002-7058-7682]{Hideko Nomura}
\affiliation{National Astronomical Observatory of Japan, 2-21-1 Osawa, Mitaka, Tokyo 181-8588, Japan}

\author[0000-0001-8798-1347]{Karin I. \"Oberg}
\affiliation{Center for Astrophysics \textbar\, Harvard \& Smithsonian, 60 Garden Street, Cambridge, MA 02138, USA}

\author[0000-0001-8642-1786]{Chunhua Qi}
\affiliation{Center for Astrophysics \textbar\, Harvard \& Smithsonian, 60 Garden Street, Cambridge, MA 02138, USA}

\author[0000-0002-6429-9457]{Kamber R. Schwarz}
\altaffiliation{NASA Hubble Fellowship Program Sagan Fellow}
\affiliation{Lunar and Planetary Laboratory, University of Arizona, 1629 E. University Blvd, Tucson, AZ 85721, USA}

\author[0000-0003-1534-5186]{Richard Teague}
\affiliation{Center for Astrophysics \textbar\, Harvard \& Smithsonian, 60 Garden Street, Cambridge, MA 02138, USA}

\author[0000-0002-6034-2892]{Takashi Tsukagoshi} 
\affiliation{National Astronomical Observatory of Japan, 2-21-1 Osawa, Mitaka, Tokyo 181-8588, Japan}

\author[0000-0003-1526-7587]{David J. Wilner}
\affiliation{Center for Astrophysics \textbar\, Harvard \& Smithsonian, 60 Garden Street, Cambridge, MA 02138, USA}

\author[0000-0003-4099-6941]{Yoshihide Yamato}
\affiliation{Department of Astronomy, Graduate School of Science, The University of Tokyo, Tokyo 113-0033, Japan}

\author[0000-0002-0661-7517]{Ke Zhang}
\altaffiliation{NASA Hubble Fellow}
\affiliation{Department of Astronomy, University of Michigan, 323 West Hall, 1085 South University Avenue, Ann Arbor, MI 48109, USA}
\affiliation{Department of Astronomy, University of Wisconsin-Madison, 475 N Charter St, Madison, WI 53706}

\begin{abstract}

During the main phase of evolution of a protoplanetary disk, accretion regulates the inner-disk properties, such as the temperature and mass distribution, and in turn, the physical conditions associated with planet formation.
The driving mechanism behind accretion remains uncertain; however, 
one promising mechanism is the removal of a fraction of angular momentum via a magnetohydrodynamic (MHD) disk wind launched from the inner tens of astronomical units of the disk.
This paper utilizes CO isotopologue emission to study the unique molecular outflow originating from the HD~163296 protoplanetary disk obtained with the Atacama Large Millimeter/submillimeter Array. 
HD~163296 is one of the most well-studied Class II disks and is proposed to host multiple gas-giant planets. 
We robustly detect the large-scale rotating outflow in the $\mathrm{^{12}CO}$ $J=2-1$ and the $\mathrm{^{13}CO}$ $J=2-1$ and $J=1-0$ transitions. 
We constrain the kinematics, the excitation temperature of the molecular gas, and the mass-loss rate.
The high ratio of the rates of ejection to accretion ($5 - 50$), together with the rotation signatures of the flow, provides solid evidence for an MHD disk wind.
We find that the angular momentum removal by the wind is sufficient to drive accretion though the inner region of the disk; therefore, accretion driven by turbulent viscosity is not required to explain HD~163296's accretion. The low temperature of the molecular wind and its overall kinematics suggest that the MHD disk wind could be perturbed and shocked by the previously observed high-velocity atomic jet. This paper is part of the MAPS special issue of the Astrophysical Journal Supplement.

\end{abstract}


\section{Introduction} \label{sec:intro}

In recent years, observations of protoplanetary disks across multiple wavelength regimes have focused on reaching smaller and smaller spatial scales \citep[e.g.][]{2018ApJ...869L..41A, 2018ApJ...863...44A,oberg20}. 
The goal of these studies is to increase our understanding of the evolution of disks and the planet formation process(es) occurring within them. It is widely accepted that disks mediate the accretion of gas and dust from the envelope onto the star \citep{doi:10.1146/annurev-astro-081915-023347}, but the driving mechanism of mass transport within the disk is still heavily debated. 
Most disk evolution and planet formation theories work within the construct of the magnetorotational instability \citep[MRI;][]{1991ApJ...376..214B}. Over the past decades it has been realized that due to the low ionization fraction in disks the MRI will be quenched in extended ($\approx$1-10~au) regions of the disk called dead zones \citep{1996ApJ...457..355G,2011ApJ...739...50B}. 
The presence of such dead zones is supported by observations and astrochemical modeling \citep{2011ApJ...743..152O,2012ApJ...747..114W,2015ApJ...799..204C}.
Here the MRI will not be effective at driving accretion \citep[see][and references therein]{2014prpl.conf..411T}.
The accretion process sets the mass distribution and influences the migration timescale and direction of forming planets in the disk \citep[e.g.][]{2018A&A...615A..63O,2020A&A...633A...4K} and thus is of key importance to constrain. 

The few informative measurements of turbulence in line-emitting layers of disks show that the inferred viscosity is at least one to two orders of magnitude too low for the MRI to be the main driving force for accretion in these layers \citep[e.g.][]{2016A&A...592A..49T,2017ApJ...843..150F}. Similar results are found from looking at the degree of dust settling in HL Tau, which is likely probing much closer to the disk midplane \citep{2016ApJ...816...25P}. This difference can be reconciled if the primary mechanism of angular momentum loss is due to a disk wind rather than the MRI \citep{2017ApJ...845...31H, 2018MNRAS.475.5059K}. It is well known that accretion can alternatively be driven by a magnetocentrifugal (or MHD) disk wind, even when the MRI is quenched as long as the magnetic field has a nonvanishing flux \citep{1982MNRAS.199..883B,1997A&A...319..340F,2007prpl.conf..277P, 2013ApJ...769...76B}. If the disk is threaded with a magnetic field, the twisting of the field lines in the upper layers of the disk atmosphere can lead to the removal of angular momentum from the disk in the vertical direction along toroidal field lines. 
MHD disk winds launched from a few astronomical units are slow ($\approx20$~km~s$^{-1}$; \citealt{2007prpl.conf..277P})
compared to the high velocity ($\approx200$~km~s$^{-1}$) jets from the inner region $\approx <2$~au \citep[see][, and references therein]{2006A&A...453..785F}. The wind dynamics, and importantly the magnitude of the angular momentum loss, depend crucially on the disk magnetization, surface heating, and ionization structure, all of which remain poorly constrained in disks \citep{2017A&A...600A..75B, 2019ApJ...874...90W}.

Observations of disk winds therefore provide indirect information about these properties. 
Observational evidence of molecular MHD disk winds in young Class 0/I systems has been growing rapidly over the past few years, primarily due to the high angular resolution and sensitivity provided by the Atacama Large Millimeter/submillimeter Array (ALMA), which are required to resolve the rotation signature of the wind.
In protostars, molecular-disk winds have been traced in CO, SO, and CS line emission
\citep[e.g.][]{2009A&A...494..147L,2016Natur.540..406B,2017A&A...607L...6T,2018ApJ...864...76Z,2017NatAs...1E.146H,2019ApJ...883....1Z,2020A&A...634L..12D}, and the launch regions for the winds range from $\approx$1-40~au.  
Thus, the wind is launched from the region of the disk where planet formation takes place. In the more evolved Class II disks, observations of atomic T Tauri jets in the optical have provided the first evidence for the presence of MHD disk winds launched within $\lesssim 3$~au \citep{2002ApJ...576..222B,2003ApJ...590L.107A,2004A&A...416L...9P}. However, ALMA observations of MHD disk wind candidates launched at larger distances remain scarce \citep{2018A&A...618A.120L}.

HD~163296 is a nearby Herbig Ae star (101.5~pc, A0, 1.95~M$_{\odot}$, $\approx6$~Myr; \citealt{2018A&A...616A...1G, 2020MNRAS.493..234W})
that is host to one of the most well-studied protoplanetary disks. 
The evidence for multiple gas-giant planets in the dust and gas observations
\citep{2016PhRvL.117y1101I, 2018ApJ...860L..12T, 2018ApJ...860L..13P}
coupled with the proximity of the source make it a unique observational laboratory for studying giant planet formation. 
Although the star is relatively old it is still actively accreting matter at a relatively high rate of 
$\mathrm{log(\dot{M}_{acc})} = -6.79^{+ 0.15}_{+ 0.15}$ M$_{\odot}$ $\mathrm{yr^{-1}}$ \citep{2020MNRAS.493..234W}. 
This accretion rate has been shown to have increased by $\approx$1 order of magnitude on a time-scale of $\approx$15~yr \citep{2013ApJ...776...44M,2014A&A...563A..87E}. 
Accretion and ejection are inherently linked and this variability in accretion rate can be related to a periodic outflow sampled by HH~409
\citep{2006ApJ...650..985W, 2014A&A...563A..87E}.
Associated with the high~velocity jet emission ($\approx 250$~km~s$^-1$) from the near side of the disk is a slow ($\approx$20 km$s^{-1}$ blue-shifted) molecular outflow that has been detected in $\mathrm{^{12}CO}$ line emission with ALMA \citep{2013A&A...555A..73K}. 
Determination of the launching mechanism and launch region of this outflow requires higher angular resolution and sensitivity observations.

In this paper, we report the characterization of the HD~163296 molecular outflow in multiple CO isotopologues. 
These data were collected during the Cycle~6 ALMA Large Program, The Chemistry of Planet Formation (see \citet{oberg20} for further details) designated with the acronym MAPS (Molecules with ALMA on Planet-forming Scales\footnote{\url{www.alma-maps.info}}).
We present high-resolution observations of $\mathrm{^{12}CO}$ $J=2-1$ emission
combined with the $\mathrm{^{13}CO}$ multiline data. This allows a first data-driven determination of the column density of material ejected and excitation temperature of the wind using both $\mathrm{^{13}CO}$ and $\mathrm{^{12}CO}$ line emission. 
We use these data to investigate the location of the launch region, the mass and angular momentum loss rates, and the dynamical timescale of the wind. 
We also discuss the properties of the wind in the larger context of the other signatures of variability in the HD~163296 system and the potential connections to, and influence on, the ongoing planet formation in the disk. 

\section{Methods} \label{sec:obs}

\begin{table*}
\caption{Observations of CO isotopologues towards the HD~163296 system}
\centering
\small
\begin{tabular}{c c c c c c c c}
\hline 
Molecule/Transition      &  Frequency  & $\mathrm{E_{up}}$ & Beam (PA) & $\delta v$ &rms  & rms$_{\mathrm{pbcorr}}$ & Peak Flux  \\ 
&     (GHz) & (K) & & (\kms) & (mJy~beam$^{-1}$) &(mJy~beam$^{-1}$) &  (mJy~beam$^{-1}$) \\ \hline \hline
$\mathrm{^{12}CO}$ $J=2-1$ & 230.538000   & 16.6   & 0\farcs30$\times$0\farcs25 (85$^{\circ}$)  & 0.2  & 1.12  & 3.05 & 12.4 \\
$\mathrm{^{12}CO}$ $J=2-1$ & 230.538000   & 16.6   & 1\farcs0$\times$0\farcs85  ($85^{\circ}$)  & 0.5  & 2.20  & 5.00 & 78.4  \\
$\mathrm{^{13}CO}$ $J=2-1$ & 220.398684   & 15.9   & 1\farcs0$\times$0\farcs85  ($66^{\circ}$)    & 0.5  & 1.70  & 4.45 & 40.2 \\
$\mathrm{^{13}CO}$ $J=1-0$ & 110.201354   & 5.3   & 1\farcs0$\times$0\farcs85  (68$^{\circ}$)    & 0.5  & 1.12  & 2.95 & 24.0  \\
\hline 
\end{tabular}
\label{table1}
\begin{tablenotes}
\item{Line frequencies and upper energy levels are taken from the Cologne Database for Molecular Spectroscopy (CDMS) \citep{2005JMoSt.742..215M}.}
\end{tablenotes}
\end{table*}

\subsection{Observations}

\begin{figure*}
    \centering
    \includegraphics[trim={1cm 0 1cm 1cm},clip, width=0.9\hsize]{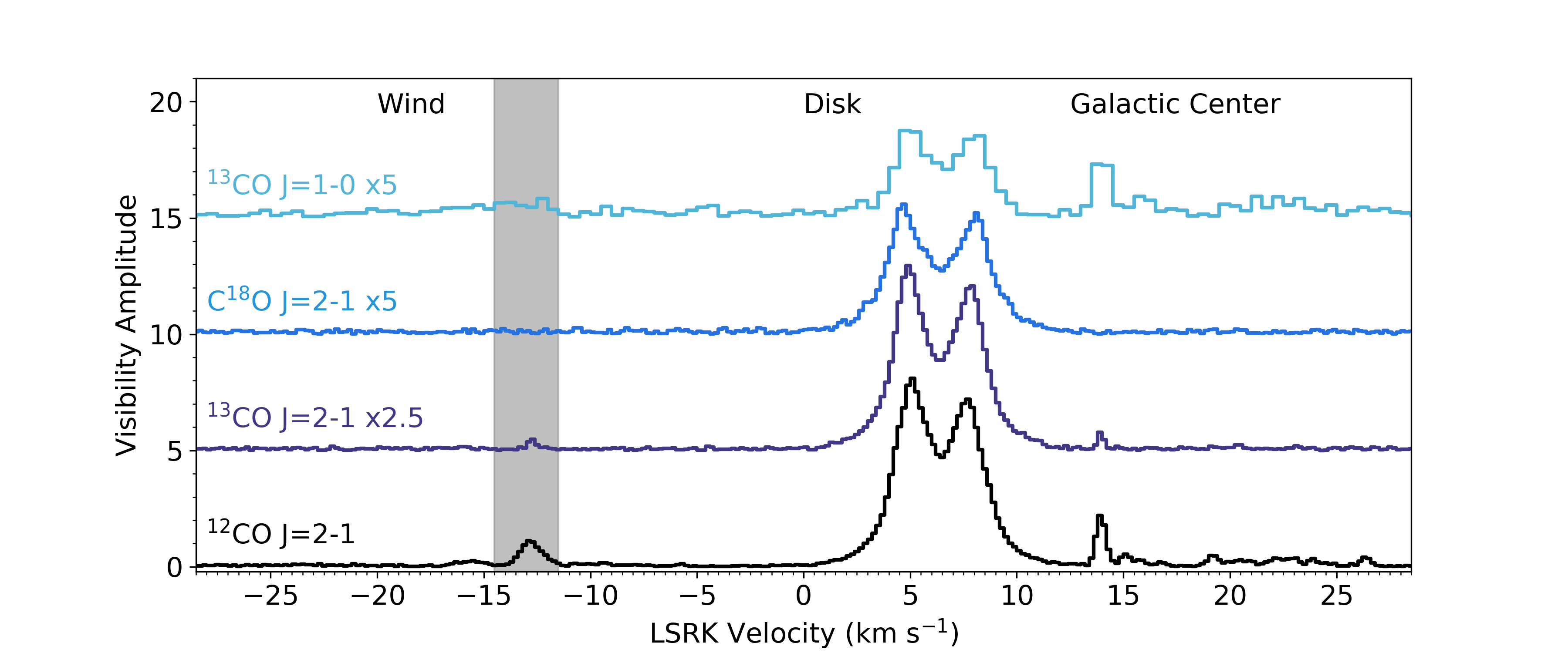}
    \caption{Average visibility spectra over the four shortest baselines for four CO isotopologues and transitions. 
    The $\mathrm{^{13}CO}$ $J=2-1$, $\mathrm{C^{18}O}$ $J=2-1$ and $\mathrm{^{13}CO}$ $J=1-0$ data are offset by values of 5, 10 ,and 15 respectively, and multiplied by a factor noted on the figure for clarity. The velocity resolution for each line is noted in Table~\ref{table1}.
    The shaded regions highlight the CO tracing the blue-shifted disk wind (left) and CO from the Galactic center (right).}
    \label{uvplots}
\end{figure*}

The data presented here were collected as part of the ALMA Large Program, “The Chemistry of Planet Formation” (2018.1.01055.L), with co-PIs, K. I. \"{O}berg, Y. Aikawa, E. A. Bergin, V. V. Guzm\'{a}n, and C. Walsh.
This paper is focused on the HD~163296 disk and the CO line data only. 
These select CO isotopologues and transitions used in this paper are listed in Table~\ref{table1}. 
For full details on the program and the data reduction process please see the overview paper \citet{oberg20}. 
We first checked for the presence of the blue-shifted wind as detected in \citet{2013A&A...555A..73K} by inspecting the data in CASA using \texttt{plotms}. 
Figure~\ref{uvplots} presents the $uv$ amplitude of the data as a function of velocity averaged over the four shortest baselines (19, 28, 29, and 32~m). 
The channel widths chosen for the $J=2-1$ and $J=1-0$ lines are 0.2~\kms and 0.5~\kms, respectively. Full information on the data calibration can be found in \citet{oberg20}.
The blue-shifted wind is most apparent in the $\mathrm{^{12}CO}$ $J=2-1$ line from $\approx-14~\mathrm{to}~-12$~\kms LSRK (highlighted in the left-hand gray box in Figure~\ref{uvplots}). 
This feature is also detected in the $\mathrm{^{13}CO}$ lines but is undetected in the $\mathrm{C^{18}O}$ lines (not shown in Figure~\ref{uvplots}). 
The emission from $\approx 13 - 15$~\kms LSRK is from a bright background cloud in the Galactic center, situated at a velocity far in the line wings of the emission from HD~163296 itself. 
HD~163296 is located at low Galactic latitude ($l=7.24^\circ$, $b=+1.49^\circ$) and the 1.2~m CO telescope survey by \citet{dame01} reveals bright $\mathrm{^{12}CO}$ $J=1-0$ emission at the velocities indicated in the MAPS data.

\subsection{Imaging}

As the line images generated using the MAPS imaging pipeline described in \citet{czekala20} cover the velocity range of the disk emission only, custom images were generated for this study. 
As a starting point, the \texttt{tCLEAN} optimized imaging parameters described in \citet{czekala20} were used. 
The shortest baseline (19~m) was excluded from the imaging because when included, it generated significant striping in the images that is characteristic of larger-scale emission not resolved by the interferometer. 
At Band~6, the primary beam is $\approx$38\farcs~in diameter and is $\approx 76$\farcs0~at Band~3. As noted in \citet{Huang21} the maximum recoverable scale at Band~6 is $\approx$12\farcs0. Excluding the 19~m baseline results in a maximum recoverable scale of $\approx$11\arcsec0. The spatial extent of the wind is $\approx$10\arcsec~therefore spatial filtering should not be a significant issue as this is less than the maximum recoverable scale.
From the \ce{^{13}CO}, the wind traced in the $J=2-1$ and $J=1-0$ transitions is approximately the same spatial extent. Since the Band 3 data has a much larger maximum recoverable scale, we do not appear to be missing more extended emission, but we cannot confirm this without shorter-baseline/single-dish data.

For Keplerian disk the position-velocity pattern of the line emission is well characterized and therefore we can use an analytical clean mask calculated using the stellar mass and position and inclination angles of the disk.
The velocity structure of the wind does not yield to a similar simple parameterization and therefore the clean mask was hand drawn.
All images were generated with a Briggs robust parameter of 0.5 \citep{briggs95}. 
As detailed thoroughly in \citet{czekala20} we applied a correction to the CLEANed channel maps.  This is done because the units of the residuals are in Jansky per dirty beam whereas the CLEAN model is in Jansky per clean beam. When the dirty beam is non-Gaussian these units are no longer equivalent. This was first outlined in  \citet{jorsater95}.
This so-called JvM correction is a rescaling of the residuals that are added to the CLEAN model in the final stage of the CLEAN pipeline.
The primary beam correction was then applied to the resulting images.

The $\mathrm{^{12}CO}$ $J=2-1$ line was imaged at a velocity resolution of 0.2~\kms, and to improve image quality and signal-to-noise, a $uv$ taper to force a 0\farcs30 beam major axis was used, resulting in a synthesized beam size of 0\farcs30$\times$0\farcs25 (85$\deg$) which is $\approx$30$\times$25~au. 
In order to have matching resolution data between the Band 6 and Band 3 $\mathrm{^{13}CO}$ $J=2-1$ and $J=1-0$ data, the lines were both imaged with the same $uv$ tapers at the same velocity resolution.
The beam size was chosen to give the best compromise between signal-to-noise and spatial resolution.
A $uv$ taper was used to force a beam major axis of 1\farcs00, which resulted in a synthesized beam size of 1\farcs$00\times$0\farcs85 (68$\deg$)  which is $\approx$100$\times$85~au and with a velocity resolution of 0.5~\kms. 
The $\mathrm{^{12}CO}$ $J=2-1$ transition was also reimaged with same $uv$ taper as the $\mathrm{^{13}CO}$ $J=2-1$ transition to allow for a direct comparison of the fluxes between emission from the two isotopologues. 
The rms (root-mean-square) noise from the line-free channels of the image cubes, before and after the primary beam correction, and the peak values of the emission for all lines, are listed in Table~\ref{table1}. 
The resulting channel maps for the $\mathrm{^{12}CO}$ $J=2-1$ are shown Figure~\ref{12co_chans1} in the Appendix.

\begin{figure*}
\centering
\includegraphics[width=0.75\hsize]{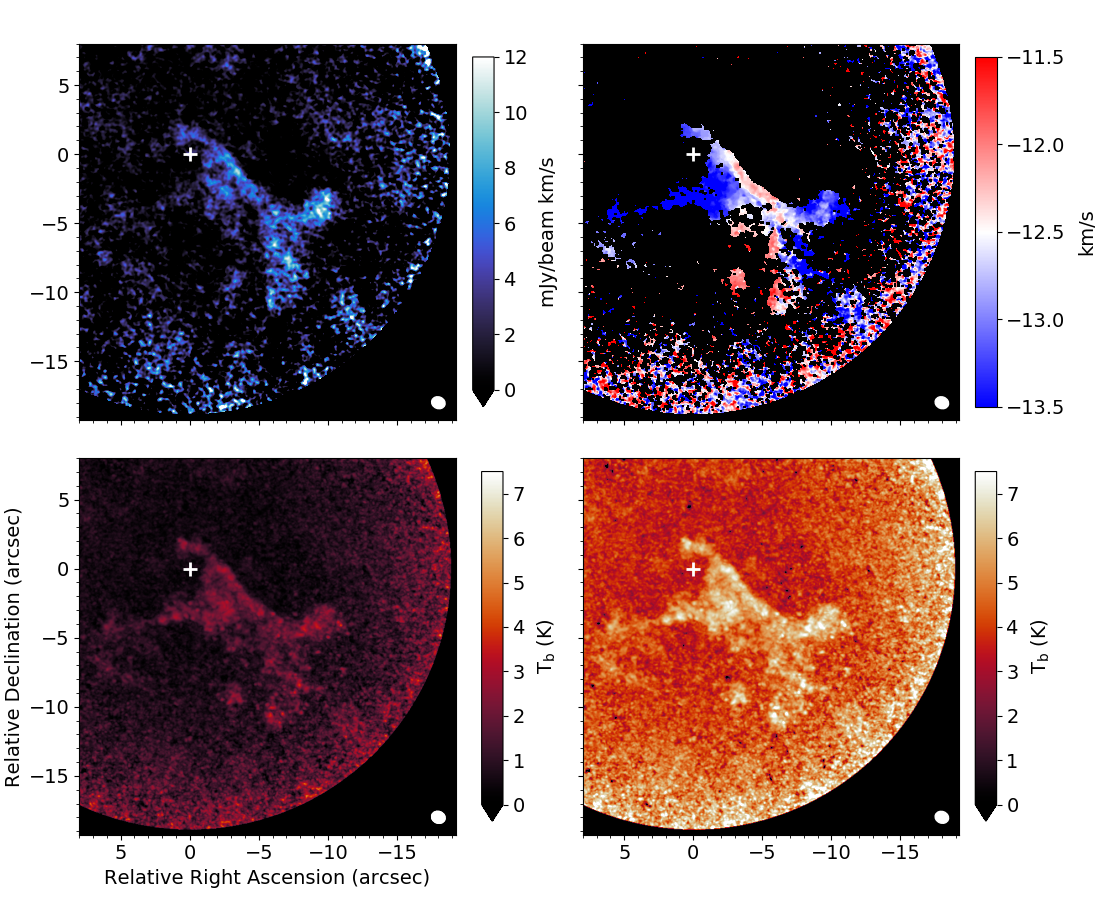}
\caption{Top left: $\mathrm{^{12}CO}$ $J=2-1$ integrated intensity map.
Top right: $\mathrm{^{12}CO}$ $J=2-1$ intensity-weighted velocity map in LSRK frame with a $4\sigma$ noise clip. 
Bottom: Peak brightness temperature maps made with the Rayleigh-Jeans approximation (left) and without (right).
In all panels the beam size is shown by the ellipse in the bottom right corner.}
\label{12co_moments}
\end{figure*}

The integrated intensity maps were made by summing over the all of the emission in channels with $>3\sigma$ emission, where $\sigma$ is the rms noise, before primary beam correction but after the JvM correction. 
For the $\mathrm{^{12}CO}$ $J=2-1$ line this covered a velocity range of -14.4 to -11.4~\kms LSRK (16~channels) and for the $\mathrm{^{13}CO}$ lines this was -13.5 to -11.5~\kms LSRK  (5~channels) for the $J=1-0$ transitions and -13.5 to -12.0~\kms LSRK  (4~channels) for the J=2-1 transition. 
Figure~\ref{12co_moments} presents the integrated intensity map and the intensity-weighted velocity map. 
The latter was generated over the same channels as the integrated intensity map but with a $4\sigma$ noise clip applied to the channel map. 
Figure~3 presents the integrated intensity maps for the $\mathrm{^{13}CO}$ $J=1-0$ and $J=2-1$ lines and the intensity-weighted velocity map.  
The latter was generated over the same number of channels as the integrated intensity map but with a $4\sigma$ noise clip applied to the channel map.
Also shown in Figures~\ref{12co_moments} and 3 are the brightness temperature maps for the $\mathrm{^{12}CO}$ and both of the $\mathrm{^{13}CO}$ lines. These were generated from the peak intensity (moment~8) maps using the \texttt{imagecube.jybeam_to_Tb_RJ} and \texttt{imagecube.jybeam_to_Tb} functions in the \texttt{gofish} package \citep{GoFish}\footnote{\url{https://github.com/richteague/gofish}}.

\section{Analysis} \label{sec:analysis}

\subsection{Emission morphology}

The integrated intensity maps presented in Figure 2 show that the width of the emission along the axis perpendicular to the jet varies. The emission close to the disk is 
$\approx$500~au wide, then this narrows to $\approx$100~au and then widens again. 
This is also seen in the channel maps shown in Figure~\ref{12co_chans1} in the Appendix. 
We recover the double corkscrew structure in the $\mathrm{^{12}CO}$ $J=2-1$ gas kinematics in the intensity weighted velocity maps as first reported by \citet{2013A&A...555A..73K}. 
The same kinematic structure is seen in the two $\mathrm{^{13}CO}$ lines (see Figure~3). 

The high-velocity outflow, HH 409, from the HD~163296 system is periodic and asymmetric \citep{2006ApJ...650..985W, 2014A&A...563A..87E}. The jet from the near side of the disk has three knots (A, A2, A3) and a bow shock (H). 
Figure~3 shows the $\mathrm{^{13}CO}$ moment maps with the well-constrained axis of the blue-shifted jet and the estimated locations of the knots at the time of our observations. 
The proper motion of the blue-shifted knots is $0.49 \pm 0.01~\mathrm{yr^{-1}}$ \citep{2006ApJ...650..985W}.
This means that the knots will have moved $\approx$~3\farcs25 since the data presented in \citet{2013A&A...555A..73K} was taken. In \citet{2013A&A...555A..73K} the $\mathrm{^{12}CO}$ $J=3-2$ emission was associated with the location of knot A3. In our data A3 is the only knot that is potentially spatially associated. 
A3 is $\approx$2\farcs0 from the brightest points in the $\mathrm{^{13}CO}$ brightness temperature maps. 
Knots A and A2 are still within the primary beam of the Band 6 data but no significant emission is detected at these locations.  
Beyond the Band 6 primary beam at 28\farcs0 there is $\mathrm{^{13}CO}$ $J=1-0$ emission associated with the bow shock H.
To determine the current distance of the bow shock ($\approx$27\farcs25), the proper motion was assumed to be the same as that for the knots.

\subsection{Kinematics}

To further investigate the kinematics of the wind we make position-velocity (PV) diagrams.
We take cuts perpendicular to the axis of the jet and average over one beam.
We present Figure~\ref{pv_example} as an example case in the main text and the rest are shown in the Appendix in Figure~\ref{fig:awesome_image}. 
The transverse velocity gradient shown in the moment~1 maps are recovered in the PV diagrams and are in the form of a coherent tilt, highlighted by the red line, that is suggestive of rotation for projected distances, $z \lesssim 600~$au. Quantitatively, the observed tilt in the PV diagram at $z=480$~au has a spatial extent and velocity range of $\Delta r \sim 4$\farcs~and $\Delta V \sim 1.6$~\kms, respectively. 
Interpreting this tilt as rotation, this gives a rotation velocity of $V_{\phi} \equiv \frac{\Delta V}{2 \sin{i}} = 1.1$~\kms and a specific angular momentum of $j \equiv V_{\phi} \Delta r/2 = 220$ km s$^{-1}$ au where $i = 46.7\deg$ is the inclination of the flow with respect to the line of sight.

\begin{figure*}
    \centering
    \includegraphics[width=0.7\hsize]{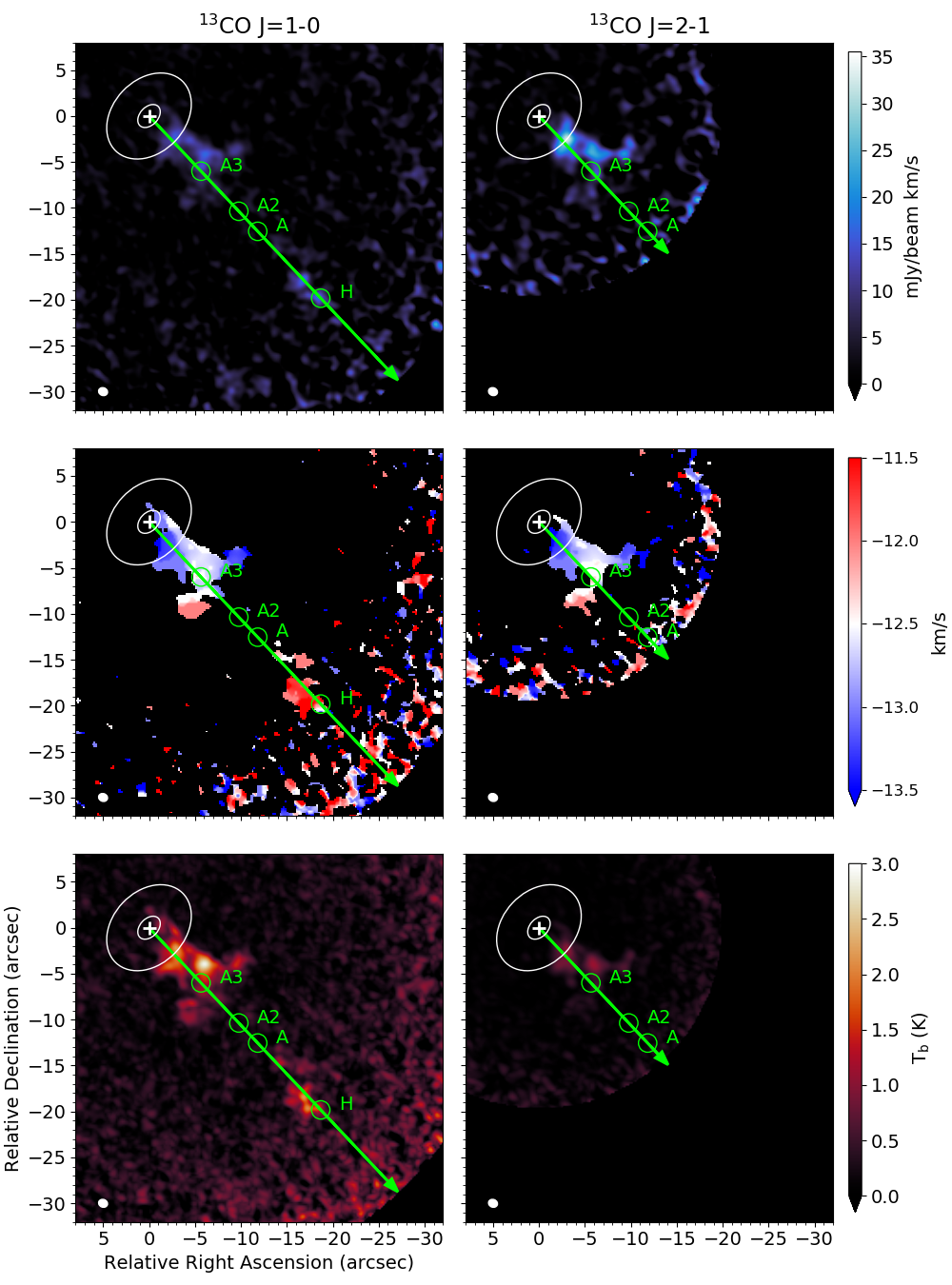}
    \caption{Integrated intensity (top), intensity-weighted velocity in LSRK frame (middle), and peak brightness temperature (bottom) maps for the $\mathrm{^{13}CO}$ $J=1-0$ and $J=2-1$ lines. 
    Ellipses are disk radii of 100 and 400~au. 
    The green arrow highlights the axis of the jet and the circles highlight the position of the knots (A, A2, A3) and the bow shock (H) with corrections for proper motions. 
    The beam size is shown by the ellipse in the bottom-left corner of each image and the star position is marked with a cross. A figure without annotations are shown in the Appendix in Figure~\ref{13co_moments_app}}
    \label{13co_moments}
\end{figure*}

\begin{figure}
    \centering
\includegraphics[width=0.95\hsize]{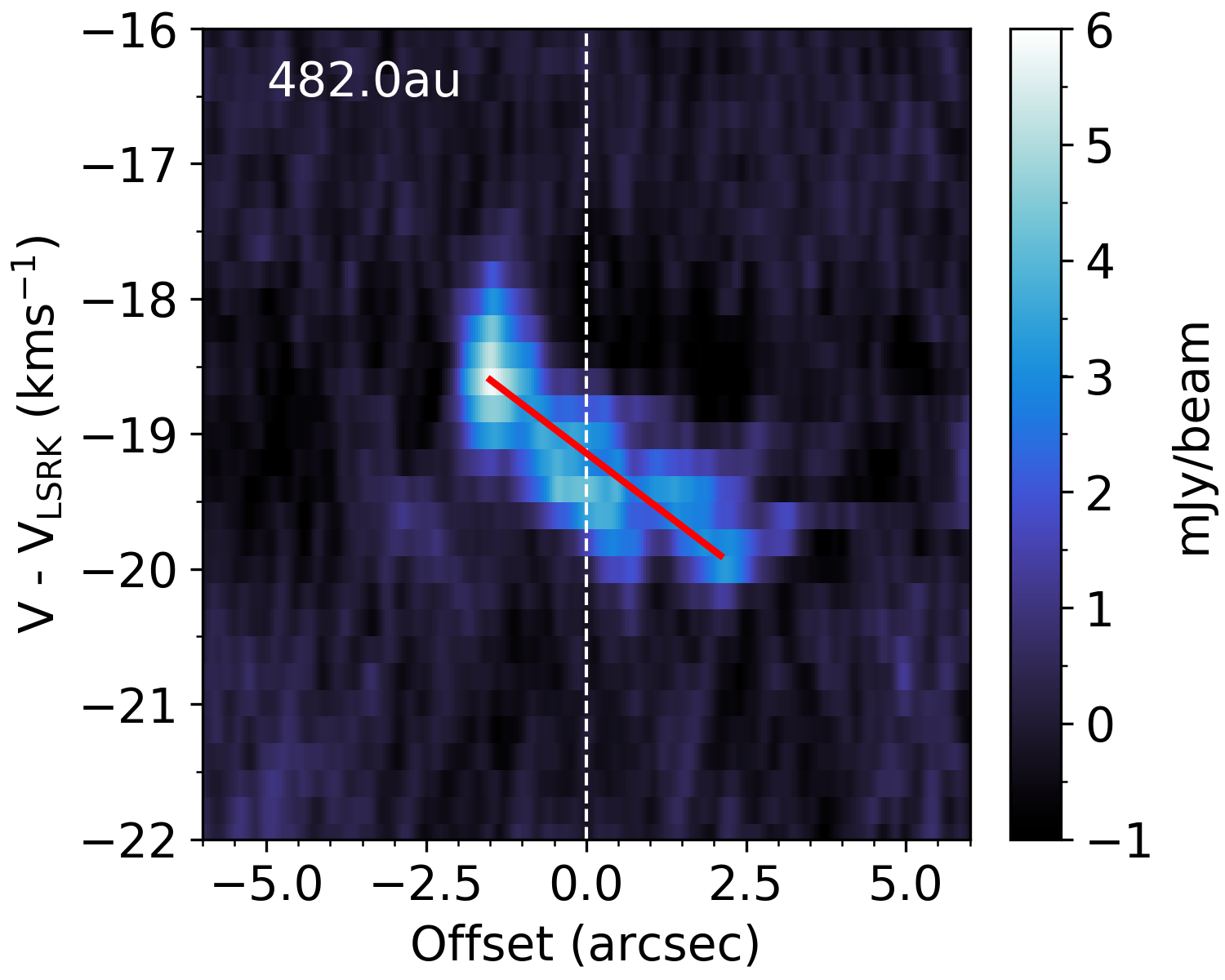}
\caption{Position-velocity diagram from a projected distance of 482~au from the star averaged over a cut that is the width of the beam parallel to the disk major axis and perpendicular to the jet axis.  The red line highlights the tilt in the diagram that can be interpreted as rotation (see Section~4.1).
 Positive offset is in the northwest and the negative in the southeast with respect to the sky. The velocity axis has been corrected for the velocity of the source.}\label{pv_example}
\end{figure}

\subsection{Column density and excitation}

From our detection of $\mathrm{^{13}CO}$, we find that the $\mathrm{^{12}CO}$ $J=2-1$ line has to be optically thick as the peak flux ratio of the $\mathrm{^{12}CO}$ $J=2-1$ and $\mathrm{^{13}CO}$ $J=2-1$ lines is $\approx 2$ in the channel maps (see Table~\ref{table1}). This result is in contrast to previous work. 
In \citet{2013A&A...555A..73K} the CO column density (\textit{N(CO)}, $\mathrm{cm^{-2}}$) and kinetic temperature ($\mathrm{T_{Kin}}$, K) of the wind were determined from the $\mathrm{^{12}CO}$ $J=3-2$ and $J=2-1$ lines. 
Their analysis focused on a single knot of emission where both lines were detected and here both lines were assumed to be optically thin with $T_{\mathrm{{Kin}}}= 960$~K and \textit{N(CO)}$=1\times10^{15}~\mathrm{cm^{-2}}$.

\begin{figure}
    \centering
    \includegraphics[width=0.9\hsize]{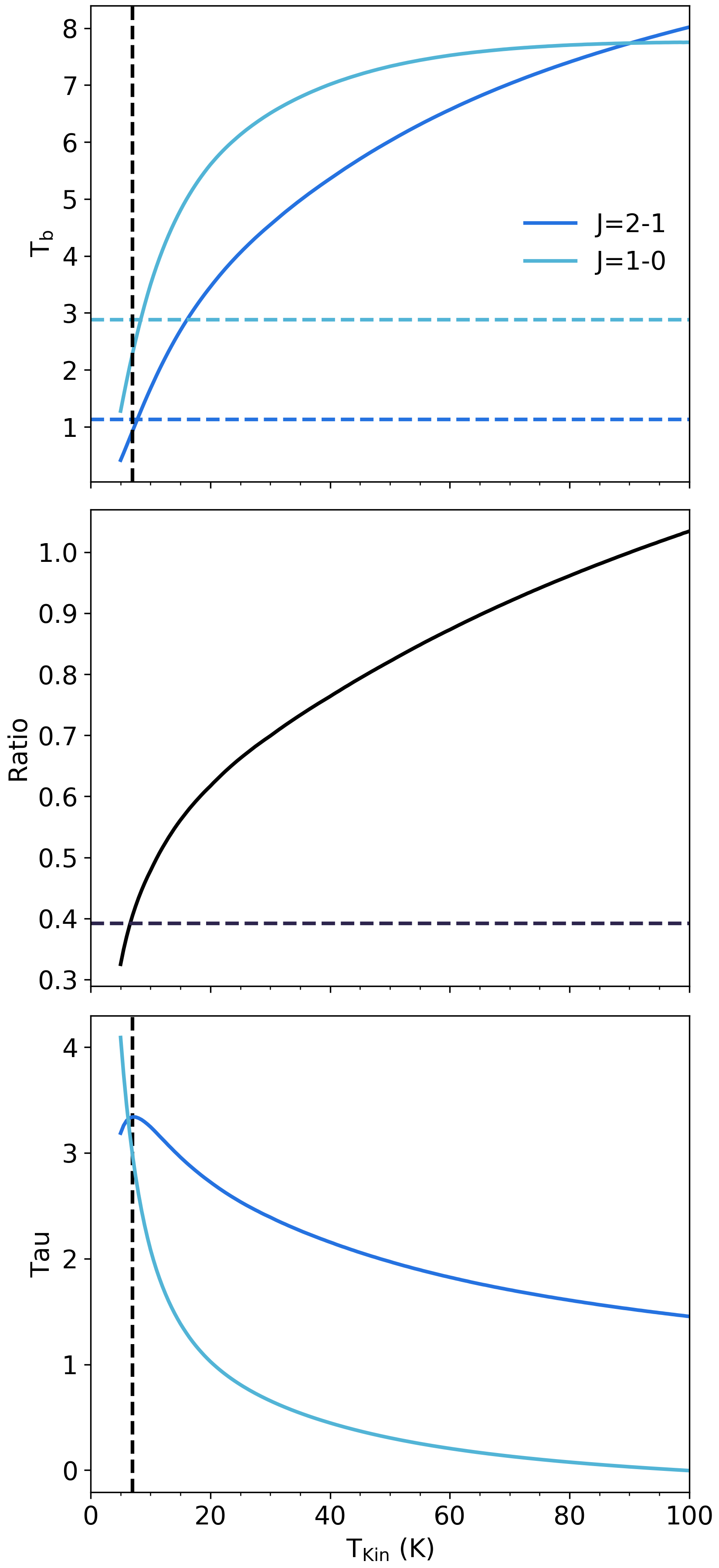}
    \caption{RADEX model results as a function of kinetic temperature 
    $\mathrm{T_{Kin}}$ 
    for the $\mathrm{^{13}CO}$ $J=1-0$ and $J=2-1$ transitions assuming a line width of 1~\kms,
    $\mathrm{n_{H}}$ of $10^3~\mathrm{cm^{-3}}$ and
      $\mathrm{\Sigma(\mathrm{^{13}CO})}$ of $4.7\times 10^{15}~\mathrm{cm^{-2}}$.
    Top: brightness temperatures, $\mathrm{T_{b}}$, from the model (solid lines), peak value 
    from the observations (dashed horizontal lines) and and best-fit kinetic temperature (dashed vertical line).
    Middle: $J=2-1/J=1-0$ ratio from the model (solid line) and observations (dashed line).
    Bottom: optical depth, $\tau$, of both lines (solid lines) and best-fit kinetic temperature (dashed line).
   }
    \label{13co_tex}
\end{figure}

Because the molecular wind is traced in both of the $\mathrm{^{13}CO}$ $J=2-1$ and $J=1-0$ lines, this ratio will provide additional constraints on the column density and excitation conditions. 
The ratio of the peak brightness temperature maps as shown in Figure~3 was calculated, resulting in a $J=2-1/J=1-0$ ratio of 0.39 where 1.13~K and 2.88~K are the peak values for each line, respectively. 
To determine the kinetic temperature, $T_{\mathrm{Kin}}$, we used the non-LTE radiative transfer code RADEX \citep{2007A&A...468..627V}.
RADEX calculates the brightness temperature, $\mathrm{T_b}$, under the Rayleigh-Jeans approximation; therefore, the output is expected to be consistent with the maps generated in Section~2.2.
We first calculate ${T_b}$ for both lines over a grid in temperature ($T_{\mathrm{Kin}}$ from 5 to 1000~K in steps of 5~K) and an assumed a gas density of $n_{\mathrm{H}} = 1.9\times10^3$ $\mathrm{cm^{-3}}$.
This density is the same as that adopted in \citet{2013A&A...555A..73K} which is based on an independent measurement of the $n_{\mathrm{H}}$ volume density in the knot \citep[1.9$\times 10^3$ $\mathrm{cm^{-3}}$;][]{2006ApJ...650..985W}.
This is not necessarily the $n_{\mathrm{H}}$ density in the wind. 
However, because of the low ($\approx~10^3$ cm$^{-3}$) critical densities of the low-lying rotational transitions of CO the hydrogen gas density is not a key parameter in the fit.
We fix the line width to 1~\kms and first use the same $\mathrm{^{12}CO}$ column density as \citet{2013A&A...555A..73K} reduced by a factor of 70 for $\mathrm{^{13}CO}$ to account for the elemental \ce{^{12}C}/\ce{^{13}C} ratio.  
Under these conditions, the brightness temperatures for the lines are a factor of 100 too low when compared to the observations. 
The results from the above RADEX models are shown in the Appendix in Figure~\ref{13co_tex_klaassen}.
The $\mathrm{^{13}CO}$ column density was gradually increased (in steps of $1\times10^{15}$~cm$^{-2}$ when the values were close) until the observed brightness temperatures were reproduced at a temperature consistent with the line ratio. 
This was done over a finer temperature grid of a smaller range ($T_{\mathrm{Kin}}$ from 5 to 100~K in steps of 0.5~K). 
This is achieved with a $\mathrm{^{13}CO}$ column density of $9.0\pm~1.0\times10^{15}~\mathrm{cm^{-2}}$ at a kinetic temperature of 7.5~K. 
The brightness temperatures, line ratio, and optical depths are presented in Figure~\ref{13co_tex}. 
This is a considerably lower temperature than \citet{2013A&A...555A..73K} who found 960 K.  However, our data clearly rule this out as the $\mathrm{^{12}CO}$ brightness temperature at peak is approximately the same as our temperature derived from the line ratios. 
The largest error on the derived column density depends on the assumed line width, e.g., with a line width of 0.5~\kms the required N($\mathrm{^{13}CO}$) to match the observations is 2 times lower than calculated with 1.0~km~s$^{-1}$. 
{An additional caveat is that we assume both the $2-1$ and $1-0$ lines have the same emitting area. This is reasonable given the data in hand, and the lines are close in excitation so should be tracing similar regions of the outflow.}

\subsection{Mass loss in the wind}

The dynamical timescale, i.e, the time since the launching event, can be estimated from the velocity of the gas and the spatial extent of the emission. 
This assumes that the flow is perpendicular to the ecliptic plane of the disk.
For the rotating wind, the deprojected length of the flow is $\approx$1400~au (assuming an inclination angle of $46.7^{\circ}$; \citealt{2018ApJ...869L..42H}). 
The deprojected velocity is 
-28~\kms, where the projected velocity is taken as the average velocity of the wind -12.9~\kms~LSRK, corrected for the velocity of the source ($+$~5.76~\kms; \citealt{2019Natur.574..378T}).
The dynamical timescale, $t_{\mathrm{dyn}}$, is therefore $\approx 240$ years. 
We can do the same calculation for the knot detected in $\mathrm{^{13}CO}$ $J=1-0$ line emission associated with bow shock H and this results in $t_{\mathrm{dyn}}\approx500$~yrs.

The mass in the wind can be estimated from the $\mathrm{^{13}CO}$ column density required in the RADEX models in Section~3.2. 
When assuming a $\mathrm{^{12}CO}$/$\mathrm{^{13}CO}$ abundance ratio of 70 the column density of $\mathrm{^{12}CO}$ in the wind is 
$6.3\times10^{17}~\mathrm{cm}^{-2}$. 
This can be considered a lower limit since isotope-selective photodissociation preferentially destroys $\mathrm{^{13}CO}$ over $\mathrm{^{12}CO}$ in the case where the $\mathrm{^{12}CO}$ is self-shielding and $\mathrm{^{13}CO}$ is not \citep{2009A&A...503..323V}. 
If active, this would increase the $\mathrm{^{12}CO}$/$\mathrm{^{13}CO}$ ratio in the material in the surface of the disk that is carried away by the wind, or indeed within the wind itself. 
This can then be converted to a total gas column by assuming a \ce{CO}/\ce{H2} ratio of $10^{-4}$ and a mean molecular mass of $2.4 m_{\mathrm{{\rm{H}}}}$, where $m_{\mathrm{{\rm{H}}}}$ is the atomic mass of hydrogen. 
This mass density ($2.54\times10^{-2}$~g~cm$^{-2}$.) can be used to calculate a total gas mass by multiplying by the projected area of the wind. 
For simplicity we take this to be a rectangle with length 1000~au and width 400~au
(measured approximately from the PV diagrams, see Figure~\ref{fig:awesome_image}).
This results in a total gas mass of $M_{w} = 1.0\times10^{-3}~\mathrm{M_{\odot}}$ (or $1 M_{\mathrm{Jup}}$). 
This is 0.07\% of the total HD~163296 disk gas mass as determined in \citet{Calahan21}.
The mass-loss rate, $\dot{M}_{w}$ of the wind can then be determined from $M_{w}/t_{dyn}$ and is 
$4.8\times10^{-6}~\mathrm{M_{\odot}~yr^{-1}}$. 
The ejection-to-accretion ratio is $f_M = \dot{M}_{w}/\dot{M}_{acc} \simeq 5-50$, with the uncertainty being driven by the error uncertainty on the measurement of the absolute accretion rate and its intrinsic variability \citep{2014A&A...563A..87E, 2020MNRAS.493..234W}. 
The ratio of the mass-loss rate from the optical jet compared to the accretion rate is much lower, $\simeq~0.01-0.1$ \citep{2014A&A...563A..87E}.
Such a large mass flux is in line with previous results obtained in the rotating molecular flows from sources of different ages and masses. 
For example, \citet{2018A&A...618A.120L}, \citet{2019ApJ...883....1Z}, and \citet{2020A&A...634L..12D} derive mass-loss rates of 45, $>10$ and 35 times larger than the mass-loss rate of the optical jets in these systems. 
Regardless of the origin of the mass-loss process, our data confirm that rotating CO outflows can extract a significant mass fraction of the accreted flow.

\section{Discussion} \label{sec:discu3}

\subsection{The nature of the HD~163296 outflow}

\subsubsection{Is the outflow entrained envelope material?}

CO outflows surrounding fast collimated jets are common at earlier stages of star formation. They have been traditionally attributed to envelope material entrained by a fast wide-angle wind or a jet launched from the inner disk \citep[$0.05-1$~au;][]{1991ApJ...370L..31S,1993A&A...278..267R,2001ApJ...557..429L,2007prpl.conf..245A}. Our observations of a massive outflow at disk scales challenges this scenario since the presence of a massive envelope is excluded. As reference, Class 0 outflows, which are suggested to trace entrained envelope material, have a similar mass-loss rates as HD~163296 but are surrounded by envelopes with a mass of about $1 M_{\mathrm{\odot}}$. Therefore, the outflow of HD163296 is most likely launched from the disk itself and be a bona fide
 "disk wind". In the following, we discuss the possible origin of such a wind.

\subsubsection{Is the outflow a photoevaporative wind
?}
Photoevaporative disk winds constitute the most compelling mechanism to disperse disks in a short period of time \citep{2014prpl.conf..475A}. In this scenario, the disk is dispersed from inside out when the accretion rate in the inner disk is about that of the mass loss rate. An argument against this being a photoevaporative wind (PE) wind is that there is no central cavity observed in this disk \citep{2020A&A...636A.116K}. If the wind were driven by photoevaporation, since we calculate a mass-loss rate that is about or higher than the current mass accretion rate, this would indicate that we should be observing the disk in its short dispersal stage. This seems highly unlikely given the typical short time scale of the dispersal of the inner disk ($\simeq 10^{5}$~yr).
 
The other argument against photoevaporation lies in the absolute value of the mass-loss rate. EUV photoevaporation models predict a scaling of the mass-loss rate with disk mass and the flux of ionizing photons \citep{2004ApJ...607..890F}. Using the disk mass of HD~163296, we deduce that a photon flux of about $\phi \simeq 10^{45}~\mathrm{s}^{-1}$ is required to account for the mass-loss rate. We do not have an estimate of $\phi$ for HD~163296 but, this is much higher than 
typically assumed in models $10^{41} - 10^{42}~\mathrm{s}^{-1}$ \citep{2009ApJ...703.1203H}.
Alternatively, if the wind is powered by X-ray photoevaporation \citep{2011MNRAS.412...13O}, an X-ray luminosity of about $L_X \simeq 10^{33} \mathrm{erg~s}^{-1}$ would be required. This is substantially higher than the  $L_X \simeq 5 \times 10^{29} \mathrm{erg~s}^{-1}$  that has been measured for HD~163296 \citep{2009A&A...494.1041G}.

We note that there is also evidence for a CO wind detected in the outer edges of the HD~163296 molecular disk \citep{2019Natur.574..378T} that may be similar to the photoevaporative flow detected from IM~Lup \citep{2017MNRAS.468L.108H}. This is very different to what is being traced in the large-scale blue-shifted CO emission discussed in this paper, and it covers a different velocity range. 

In other words, the mass-loss rate measured in the CO outflow is in tension with a photoevaporative origin as extreme values of the UV photon flux or X-ray luminosity would then be required. Moreover, this would imply that HD~163296 is in the dispersal phase right before the opening of the gap. Given the probability, this is very unlikely situation.

\subsubsection{Is the Ouflow an MHD Disk Wind?}

The MHD disk-wind model is the most compelling scenario to account for the high ejection-to-accretion mass ratio and the kinematic structure of the CO outflow. Recent global numerical simulations of weakly ionized disks, including the heating of the disk atmosphere, predicts mass-loss rates of about, or larger than, the stellar accretion rate; note that this is lower than our estimates \citep{2017A&A...600A..75B,2017ApJ...845...75B,2019ApJ...874...90W}.

The radial region of the disk from which the wind originates can be determined from the specific angular momentum and the axial velocity estimated in the wind. Assuming a steady, axisymmetric, and dynamically cold (negligible enthalpy) MHD disk wind, the launch radius $\varpi_{0}$, can be deduced from the \citet{2003ApJ...590L.107A} formula

\begin{equation}
\begin{aligned}
\varpi_{0} \approx & 13 \mathrm{au}\left(\frac{\varpi_{\infty}}{100~ \mathrm{au}}\right)^{2 / 3}\left(\frac{v_{\phi, \infty}}{5~ \mathrm{km} \mathrm{s}^{-1}}\right)^{2 / 3} \\
& \times\left(\frac{v_{p, \infty}}{30~\mathrm{km} \mathrm{s}^{-1}}\right)^{-4 /3}\left(\frac{\mathrm{M_{*}}}{1 \mathrm{M_{\odot}}}\right)^{1 / 3},
\label{eq:anderson}
\end{aligned}
\end{equation}

where $\varpi_{\infty}$ is the radius of the rotating wind, $v_{\phi, \infty}$ is the rotation velocity, 
$v_{p, \infty}$ is the poloidal flow speed, and $\mathrm{M_{*}}$ is the 
mass of the central star which is taken as 2.0~$\mathrm{M_{\odot}}$ 
\citep[e.g.][]{2018ApJ...869L..49I}. From the specific angular momentum
and the axial velocity estimated in Sec. 3.2, we deduce 
$\varpi_{0} = 4~$au. This is a representative radius because, as we discuss in Section 4.2 the wind may be perturbed by the jet. This shows that the wind is coming from the inner $\approx$10~au region of the disk. This result is in line with other studies of rotating outflows that find a launching radius between 1 and 40~au \citep[e.g.][]{2009A&A...494..147L, 2016Natur.540..406B, 2017A&A...607L...6T, 2018ApJ...864...76Z, 2017NatAs...1E.146H, 2018A&A...618A.120L, 2020A&A...634L..12D}.

Another key parameter of the MHD disk wind is the magnetic lever arm parameter $\lambda$. 
This is the ratio between the specific angular momentum in the wind to that at the launching point, 

\begin{equation}
    \lambda = \frac{\varpi_{\infty} v_{\phi}}{v_K(\varpi_{0}) \varpi_{0}},
\end{equation}
where $v_K(\varpi_{0})$ is the Keplerian velocity at the launch point (we neglect here the angular momentum stored in the form of magnetic torsion, see \citet{2006A&A...453..785F}). $\lambda$ can be considered as the efficiency of the wind to drive accretion: the higher the value of $\lambda$, the less mass is required to be launched to extract a given amount of angular momentum.
The deduced magnetic lever arm parameter at $z=480$~au is $\lambda=2.3$. Therefore, the gas would increase its specific angular momentum by about a factor two in the wind via magnetic acceleration. Our estimated value lies between than that of DG~Tau~B and HH~30 \citep[$\lambda \simeq 1.6$, Class I and II objects respectivly][]{2018A&A...618A.120L,2020A&A...634L..12D} and that derived in HH~212 \citep[$\lambda = 5.5$ Class 0][]{2017A&A...607L...6T}. These relatively low values of $\lambda$, together with the high mass loss rate is well in line with recent numerical simulations that include the heating of the disk and a relatively low magnetization of the disk (plasma parameter $\beta \sim 10^3-10^5$). 
Interestingly the $\lambda$ parameter for the jet is constrained to $\leq 14$ by \citet{2014A&A...563A..87E} and on small scales ($<$~0.13~au) Br$\gamma$ emission arises from a hundreds of kilometers per second potential MHD-driven wind \citep{2015A&A...576A..84G} that has been modeled with $\lambda=3$.
This shows a change in efficiency of angular momentum extraction from the very inner regions of the system out to a few astronomical units.

\subsection{Connection to the HH~409 outflow}

\begin{figure*}[t!]
    \centering
    \includegraphics[width=0.9\hsize]{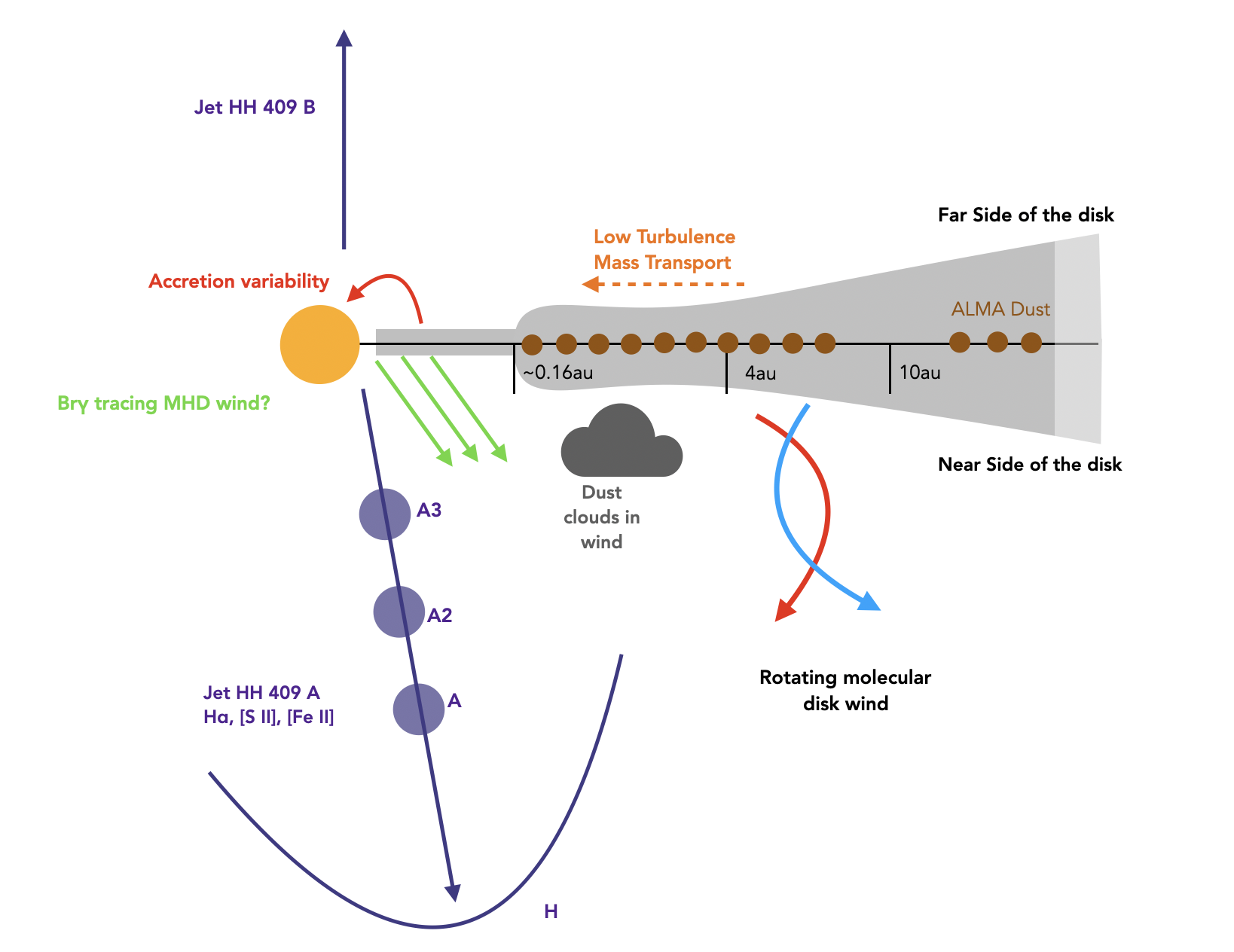}
    \caption{Cartoon summarizing the different mass-loss processes in the HD~163296 system. Note the ALMA observations presented here have a spatial resolution of 0\farcs3 ($\approx$30~au). These processes are not to scale and are for illustrative processes only.}
    \label{cartoon}
\end{figure*}

The presence of a knotty optical jet recently imaged by Very Large Telescope’s Multi Unit Spectroscopic Explorer (VLT-MUSE), the kinematical properties of the outflow, and the low temperature suggest that the pulsating jet may be interacting and compressing the wind. Figure~\ref{12co_chans3} shows the $\mathrm{^{12}CO}$ $J=2-1$ channel maps with the recent observations of 
[SII] 673nm and H$\alpha$ from \citet{2020A&A...644A.149X} overlaid. 
 
First of all, the transverse PV diagrams of CO exhibit a bar-like morphology with a straight velocity gradient (red line, Fig. \ref{pv_example}). 
This contrasts with synthetic PV diagrams of stationary MHD disk winds that show slower material off the jet axis and faster material closer to the axis \citep{2004A&A...416L...9P,2020A&A...640A..82T}. An extended MHD disk wind is indeed made of nested streamlines with axial and rotation velocities roughly scaling with the Keplerian velocity at their footpoints. The inner collimated streamlines are therefore faster than the outer wide-angle streamlines.
Therefore, the CO emission in HD~163296 traces a rotating thin shell of gas with a limited velocity shear rather than onion-like winds as expected by a stationary MHD-wind model. 
Such spatio-kinematical patterns have already been  unveiled by ALMA toward the T~Tauri outflow HH~30 \citep{2018A&A...618A.120L} and the massive Orion I Source \citep{2017NatAs...1E.146H}.

The relatively low temperature of the outflow ($\approx 7$~K) is also in tension with MHD disk-wind models. \citet{2012A&A...538A...2P} show that ion-neural friction during the MHD acceleration can indeed heat up the gas launched from a few astronimical units up to few thousand kelvin for high values of $\lambda$ ($\sim~10$) values. \citet{2019ApJ...874...90W} computed models that are more in line with the properties of the HD~163296 wind ($\lambda \approx2$) and find that temperatures of about a hundred Kelvin, an order of magnitude larger than the value derived in HD~163296.

We suggest that the appearance of the outflow as a thin shell of cold gas points toward the presence of shocks in the disk wind. These shocks could be driven by the fast pulsating jet of HD~162396 embedded in the CO outflow. In fact, the interaction between a fast pulsating jet and a slower disk wind results in the formation of a thin shell of gas swept up by the jet bow shocks \citep{2018A&A...614A.119T}. 
Most striking, in Figure~\ref{12co_chans3} over velocity channels from -12.8 to -11.80 \kms LSRK the wind is spatially offset but running parallel to the optical jet. 
Interestingly, our CO maps, as well as the CO ($J=3-2$) maps presented in \citet{2013A&A...555A..73K} do show these bow-shock structures associated with the jet. The low temperature of CO would then be the result of the compression of the wind by the shock. Further modeling is required to determine under what conditions (density, shock velocity, magnetization), the gas can cool down to $\sim 7$~K in such a short period of time $< t_{dyn} = 500$~yr.

One should then keep in mind that if the slow wind is perturbed by the fast jet, the \citet{2003ApJ...590L.107A} formula used above to derive the launching radius is not strictly applicable \citep{2016ApJ...832..152D}. MHD simulations including a pulsating jet are needed to study the biases in the derivation of the launching radius from the observed rotation signatures.

The high-velocity jet is dust free, but it is possible that the observed near infrared excess and optical variability from HD~163296 is due to dust clouds entrained in the molecular wind that are launched at the same time as the jet ejection events as discussed in \citet{2014A&A...563A..87E} and \citet{2020ApJ...902....4R}. 
Dust, in particular $<0.1~\mu$m sized grains, can also can be transported in MHD winds from the inner regions and then can fall back onto the disk at larger radii  \citep{2019ApJ...882...33G}.
This means that there is potentially mass loss of both gas and dust from the inner planet-forming zone of the disk. 
The dusty wind can also shield the disk material from far-UV radiation  \citep[e.g.][]{2012A&A...538A...2P}. This could affect the UV-driven chemistry in the upper layers of the disk.

\subsection{Does the wind drive accretion?}
The relative impact of the wind on disk accretion can be estimated by computing the fraction of disk angular momentum extracted vertically by the MHD disk wind, $f_J$, across the launching region of the wind \citep[see Eq. (36) from][]{2020A&A...640A..82T},
\begin{equation}
    f_J = \frac{\lambda}{1+\frac{ln(\varpi_{out}/\varpi_{in})}{2 ln(1+f_M)}},
\end{equation}
where $\varpi_{in}$ and $\varpi_{out}$ are the inner and outer radius of the launch region, and $f_M$ is the ratio of the mass-loss rate of the wind over the mass accretion rate onto the star, and $\lambda$ is the magnetic lever arm parameter.
In principle, $f_J$ is between 0 and 1, where 1 means all the accretion though the disk is driven by the MHD disk wind. 
One of the main caveats in the application of this formula to the flow of HD~163296 is the radial extent of the wind-launching region $\varpi_{out}/\varpi_{in}$. 
As mentioned above, the flow appears as a thin shell of gas, suggesting a narrow launching region. 
However, an extended disk wind might lose its onion-like structure if different streamlines are mixed during the interaction of the wind with the jet's bow shock.
In other words, the radial extent of the launching region remains highly uncertain and might be larger if the wind has been perturbed by the jet. 
For a relatively high value for $\varpi_{out}/\varpi_{in} = 10$, the fraction is $f_J > 1.0$. 
A narrower wind-launching region would increase $f_J$. 
Thus, the angular momentum flux extracted by the wind is at least enough required to sustain accretion through an extended disk region around $r \simeq 4$~au.
We conclude that turbulence is not necessarily required to account for accretion across this specific region of the disk, which is in line with the low value for $\alpha$ turbulence measured in the line-emitting regions, and modeled across the disk \citep{2015ApJ...813...99F, 2017ApJ...843..150F,2018ApJ...857...87L}.

\subsection{Connecting to planet formation in the inner disk}

We find that the launch radius of the HD~163296 disk wind is $\approx 4$~au, thus giving key insight into the physical process occurring in the disk on spatial scales not probed with other MAPS data. 
The cartoon in Figure~\ref{cartoon} summarizes all the processes occurring in the inner 10~au's of the disk. 
Because the wind transports mass in the disk both inwards (accretion) and vertically outwards (wind mass-loss rate), disk models with MHD-driven accretion show significantly different, much flatter, surface density profiles in the inner disk than typical MRI turbulence models \citep[e.g.][]{2016ApJ...821...80B, 2016A&A...596A..74S}.
Additionally, low turbulence will reduce the level of vertical mixing in the disk. This will impact the amount of volatile sequestration in this region of the disk 
and as a result will have important effects on the C/O ratio of gas and ice available to be accreted by forming planets here
\citep[e.g.][]{2011ApJS..196...25S,2016ApJ...833..285K}.

If the current mass -oss rate of the HD~163296 system is sustained (mass accretion rate and disk wind mass loss rate), then it will only take $\approx$0.02~Myrs for the disk to be drained of mass. This is $\approx$40$\times$ quicker than when just considering mass loss via accretion. In reality this process may take longer due to trapping of disk material in the dust rings or a decline of the mass-loss rate, but nevertheless the presence of a disk wind reduces the lifetime of the gas disk and thus the time available for giant planets to accrete their atmospheres. Although CO gas is now detected in a few debris disks the gas masses are very low, on the order of a lunar mass, and therefore would not contribute much to the atmosphere of a giant planet \citep[e.g.][]{2018ARA&A..56..541H}.

MHD-wind-driven disk evolution has been shown to alter the migration direction and timescale of planets in gas-rich disks when compared to viscous accretion models. 
Type 1 migration is suppressed with the migration timescale increasing to 1~Myr in the region where the wind is active \citep{2018A&A...615A..63O}. 
This mitigates the problem of rapid inward migration and loss of $\sim$~Mars-sized bodies before the dissipation of the gas disk. 
\citet{2020A&A...633A...4K} show that models of disks with MHD-wind-driven accretion can also lead to Type III outward migration for Saturn-to-Jupiter mass bodies. Some of their models have a $\lambda$ of 2.25, similar to what we derive for HD~163296.
In their models they have a parameter called b that is akin to a viscosity parameter.
Using our HD~163296 model parameters from \citet{Calahan21} and equations 4 and 6 in \citet{2020A&A...633A...4K}, we find that within $< 50$~au, our model has a value of b is between their $10^{-4}$ and $10^{-3}$ models.  With $b = 10^{-4}$ inward migration is seen for both Saturn- and Jupiter-mass planets, and in the case of $10^{-3}$ this leads to outward migration of Saturn-mass planets and periodic inwards and outward migration of Jupiter-mass planets.

The periodic ejection events traced by the knots in the jet bring into question the stability of the mass reservoir in the inner disk. The different mechanisms powering the ejection of material in the innermost disk via the HH~409 outflow have been heavily discussed in the literature \citep[see][and references therein]{2014A&A...563A..87E, 2020ApJ...902....4R}. 
One option is the interaction of disk material with a planet. 
The $\approx 16$~year periodicity of the ejection events corresponds to a Keplerian orbital radius of $\approx 8$~au, which is close to the position of a gap in millimeter dust at 10~au \citep{2018ApJ...869L..42H}.
Another is that the accretion variability, and also the variability in the ejected material traced via the optical jet, is due to gravitational instabilities in the inner disk \citep[e.g.][]{2009ApJ...704..715V}.
Regardless of the mechanism behind the ejection events, 
the HD~163296 star is still actively accreting and the disk evolution continues in parallel with the presence of already formed (or forming) planets \citep{2018ApJ...860L..12T, 2018ApJ...860L..13P}. 
The inner mass reservoir where planet formation is expected to be most efficient is not stable and 
the mass-loss rate of the wind is high and potential dispersal timescale quick. This will have significant effects on the formation and evolution of planets in the system.

\section{Conclusions} \label{sec:concl}

We present ALMA observations of CO isotopologues tracing the HD~163296 large-scale disk wind. 
We list here our main conclusions from our analyses of these data.

\begin{itemize}
    \item The robust detection of the $\mathrm{^{13}CO}$ $J=2-1$ line reveals that the $\mathrm{^{12}CO}$ $J=2-1$ line is optically thick in the wind. 
    \item From the $\mathrm{^{12}CO}$ $J=2-1$ brightness temperature and the $\mathrm{^{13}CO}$ $J=2-1/J=1-0$ line ratio, we show that the wind is cold, in contrast to previous analysis, with a $T_{\mathrm{Kin}}$ of 7~K and a mass-loss rate of $4.8\times10^{-6}~\mathrm{M_{\odot}~yr^{-1}}$. 
    Comparing with predictions from models, we conclude that the derived wind properties are consistent with an MHD-driven disk wind. 
    \item Interpreting the position velocity diagrams of the wind as tracing rotation results in a launch radius of 4~au. 
    The wind has a narrow shell structure that may be due to the interaction of the slow ($\approx$25 \kms) wind with the high-velocity jet ($\approx$200 \kms). 
    \item The efficiency of the angular momentum extraction by the wind is characterized by the magnetic lever arm parameter. We find a low value of $\approx$2.3 and combining this with the mass-loss rate of the wind we find that the wind removes sufficient angular momentum from the disk to drive the current accretion rate. 
    This means that the inner-disk region requires no additional source for accretion, such as that driven by turbulent viscosity.  
    \item The low temperature of the wind and the spatial offset off the wind with respect to the outflow strongly suggests interaction between the jet and the wind. This could invalidate the launch radius derived using the \citet{2003ApJ...590L.107A} formulism. Numerical simulations of disk winds, jets, and magnetic fields are required to investigate this effect thoroughly.
    \item From our determination of the wind properties, we have gained key insight into disk physics in the inner planet-forming zone ($<$10~au) of the disk. The level of turbulence, degree of vertical mixing, and the variation in the radial mass distribution will have a significant impact on the disk thermal and chemical structure and thus on the planets currently forming in the disk. 
    \item Follow-up very high spatial and spectral resolution observations with ALMA are required to probe the transition region between CO in the upper layers of the disk atmosphere and the wind. Determining the height that the disk is launched from will provide important constraint for MHD driven disk wind models and, in particular, calculating the thermal structure of the wind. 
\end{itemize}

\acknowledgments

This paper makes use of the following ALMA data: ADS/JAO.ALMA\#2018.1.01055.L. ALMA is a partnership of ESO (representing its member states), NSF (USA) and NINS (Japan), together with NRC (Canada), MOST and ASIAA (Taiwan), and KASI (Republic of Korea), in cooperation with the Republic of Chile. The Joint ALMA Observatory is operated by ESO, AUI/NRAO and NAOJ. The National Radio Astronomy Observatory is a facility of the National Science Foundation operated under cooperative agreement by Associated Universities, Inc.

A.S.B acknowledges the studentship funded by the
Science and Technology Facilities Council of the United
Kingdom (STFC) and thanks Dr. Evgenia Koumpia for some very helpful discussions on using RADEX. 
B.T. acknowledges support from the research program Dutch Astrochemistry Network II with project number 614.001.751, which is (partly) financed by the Dutch Research Council (NWO).
J.D.I. acknowledges support from the Science and Technology Facilities Council of the United Kingdom (STFC) under ST/T000287/1. 
C.W. acknowledges financial support from the University of Leeds, STFC and UKRI (grant numbers ST/R000549/1, ST/T000287/1, MR/T040726/1). 
Y.A. acknowledges support by NAOJ ALMA Scientific Research grant code 2019-13B, Grant-in-Aid for Scientific Research (S) 18H05222, and Grant-in-Aid for Transformative Research Areas (A) 20H05844 and 20H05847.
S.M.A. and J.H. acknowledge funding support from the National Aeronautics and Space Administration under Grant No. 17-XRP17 2-0012 issued through the Exoplanets Research Program. 
J.B. acknowledges support by NASA through the NASA Hubble Fellowship grant \#HST-HF2-51427.001-A awarded  by  the  Space  Telescope  Science  Institute,  which  is  operated  by  the  Association  of  Universities  for  Research  in  Astronomy, Incorporated, under NASA contract NAS5-26555.
E.A.B. and A.D.B. acknowledge support from NSF AAG grant \#1907653.
J.B.B. acknowledges support from NASA through the NASA Hubble Fellowship grant \#HST-HF2-51429.001-A, awarded by the Space Telescope Science Institute, which is operated by the Association of Universities for Research in Astronomy, Inc., for NASA, under contract NAS5-26555. 
J.K.C. acknowledges support from the National Science Foundation Graduate Research Fellowship under Grant No. DGE 1256260 and the National Aeronautics and Space Administration FINESST grant, under Grant no. 80NSSC19K1534.
G.C. is supported by NAOJ ALMA Scientific Research grant code 2019-13B. 
L.I.C. gratefully acknowledges support from the David and Lucille Packard Foundation and Johnson \& Johnson's WiSTEM2D Program. 
I.C. was supported by NASA through the NASA Hubble Fellowship grant HST-HF2-51405.001-A awarded by the Space Telescope Science Institute, which is operated by the Association of Universities for Research in Astronomy, Inc., for NASA, under contract NAS5-26555. 
V.V.G. acknowledges support from FONDECYT Iniciaci\'on 11180904 and ANID project Basal AFB-170002.
J. H. acknowledges support for this work provided by NASA through the NASA Hubble Fellowship grant \#HST-HF2-51460.001-A awarded by the Space Telescope Science Institute, which is operated by the Association of Universities for Research in Astronomy, Inc., for NASA, under contract NAS5-26555. 
C.J.L. acknowledges funding from the National Science Foundation Graduate Research Fellowship under Grant DGE1745303. 
R.L.G. acknowledges support from a CNES fellowship grant.
F.L. and R.T. acknowledge support from the Smithsonian Institution as a Submillimeter Array (SMA) Fellow.
F.M. acknowledges support from ANR of France under contract ANR-16-CE31-0013 (Planet-Forming Disks) and ANR-15-IDEX-02 (through CDP ``Origins of Life").
H.N. acknowledges support by NAOJ ALMA Scientific Research grant code 2018-10B and Grant-in-Aid for Scientific Research 18H05441.

K.I.\"O. acknowledges support from the Simons Foundation (SCOL \#321183) and an NSF AAG grant (\#1907653). 
K.R.S. acknowledges the support of NASA through Hubble Fellowship Program grant HST-HF2-51419.001, awarded by the Space Telescope Science Institute,which is operated by the Association of Universities for Research in Astronomy, Inc., for NASA, under contract NAS5-26555. 
T.T. is supported by JSPS KAKENHI Grant Numbers JP17K14244 and JP20K04017. 
Y.Y. is supported by IGPEES, WINGS Program, the University of Tokyo. 

K.Z. acknowledges the support of the Office of the Vice Chancellor for Research and Graduate Education at the University of Wisconsin – Madison with funding from the Wisconsin Alumni Research Foundation, and support of the NASA through Hubble Fellowship grant HST-HF2-51401.001. awarded by the Space Telescope Science Institute, which is operated by the Association of Universities for Research in Astronomy, Inc., for NASA, under contract NAS5-26555.

\facilities{ALMA}

\software{gofish \citep{GoFish}, CASA \citep{2007ASPC..376..127M}, RADEX \citep{2007A&A...468..627V}.}

\appendix

\section{$\mathrm{^{12}CO}$ $J=2-1$ channel maps}

\begin{figure*}
\centering
\includegraphics[trim={1cm 6cm 1.0cm 8cm},clip, width=0.7\hsize]{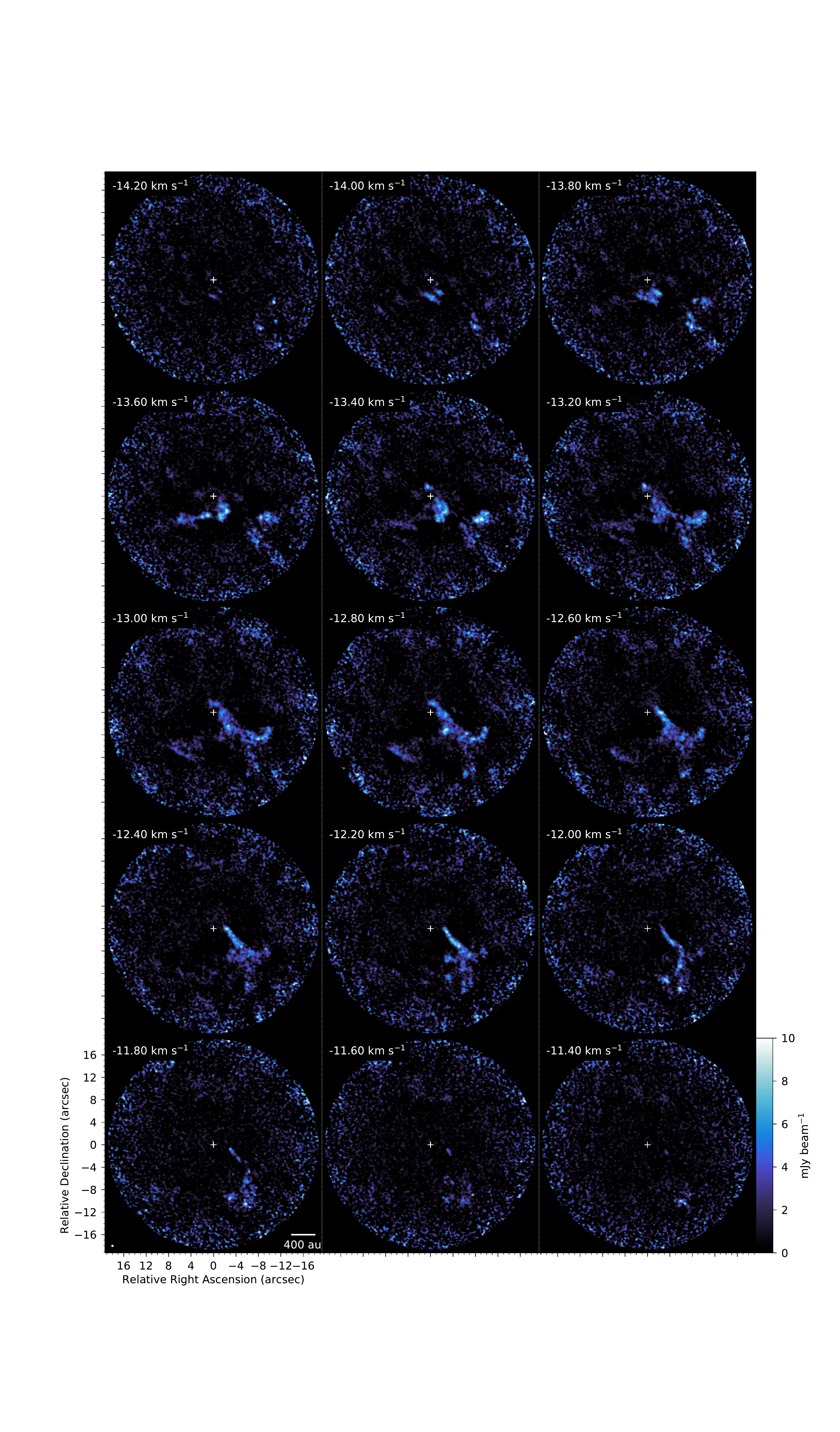}
\caption{$\mathrm{^{12}CO}$ J=2-1 channel maps. Velocity axis is LSRK frame.}
\label{12co_chans1}
\end{figure*}

\section{$\mathrm{^{12}CO}$ $J=2-1$ channel maps with annotations}

\begin{figure*}
\centering
\includegraphics[trim={1cm 6cm 1.0cm 8cm},clip, width=0.7\hsize]{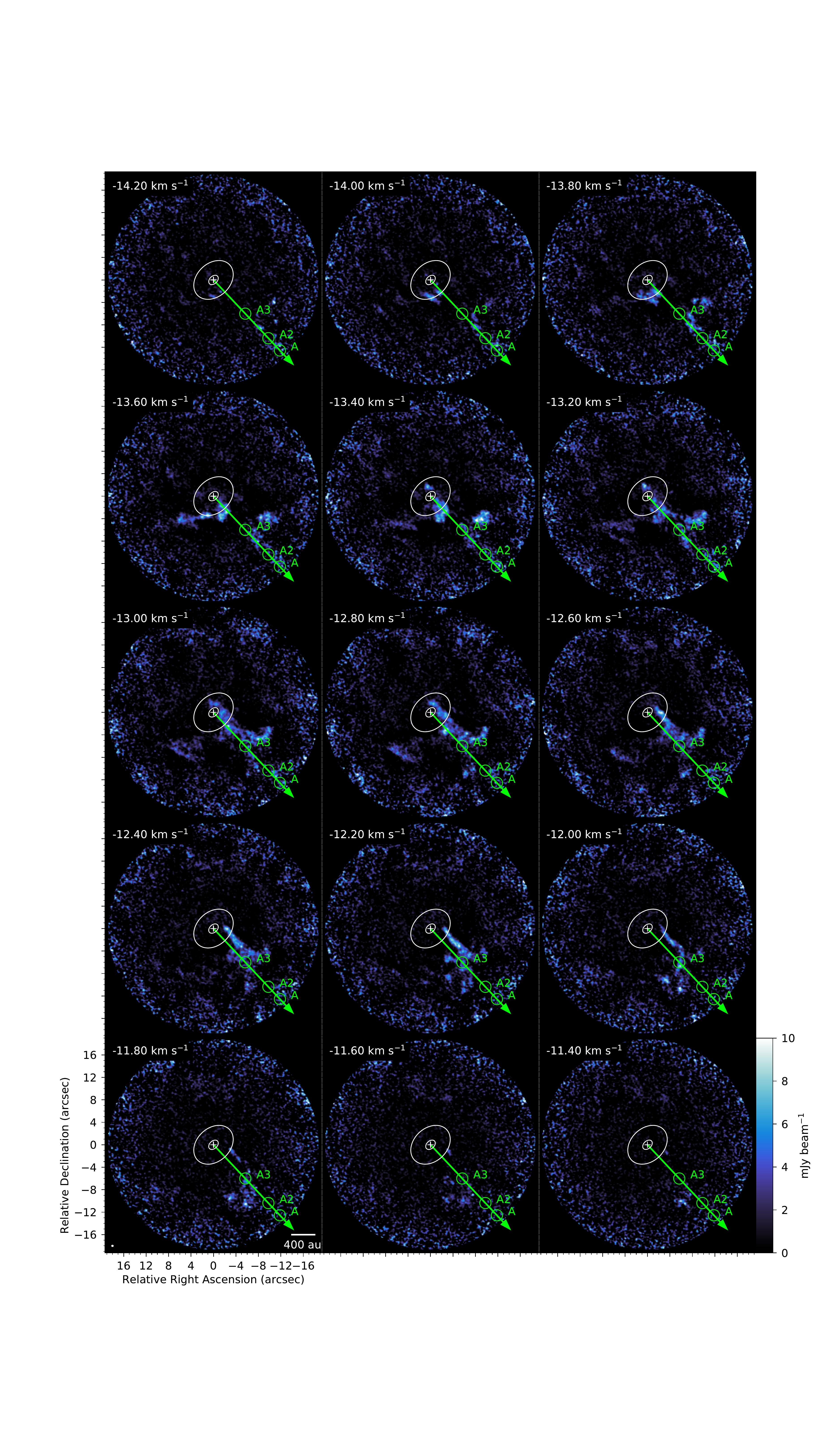}
\caption{$\mathrm{^{12}CO}$ J=2-1 channel maps with jet axis and knots labeled in green. Velocity axis is LSRK frame.}
\label{12co_chans2}
\end{figure*}

\section{$\mathrm{^{12}CO}$ $J=2-1$ channel maps with optical jet}

\begin{figure*}
\centering
\includegraphics[trim={1cm 6cm 1.0cm 8cm},clip, width=0.7\hsize]{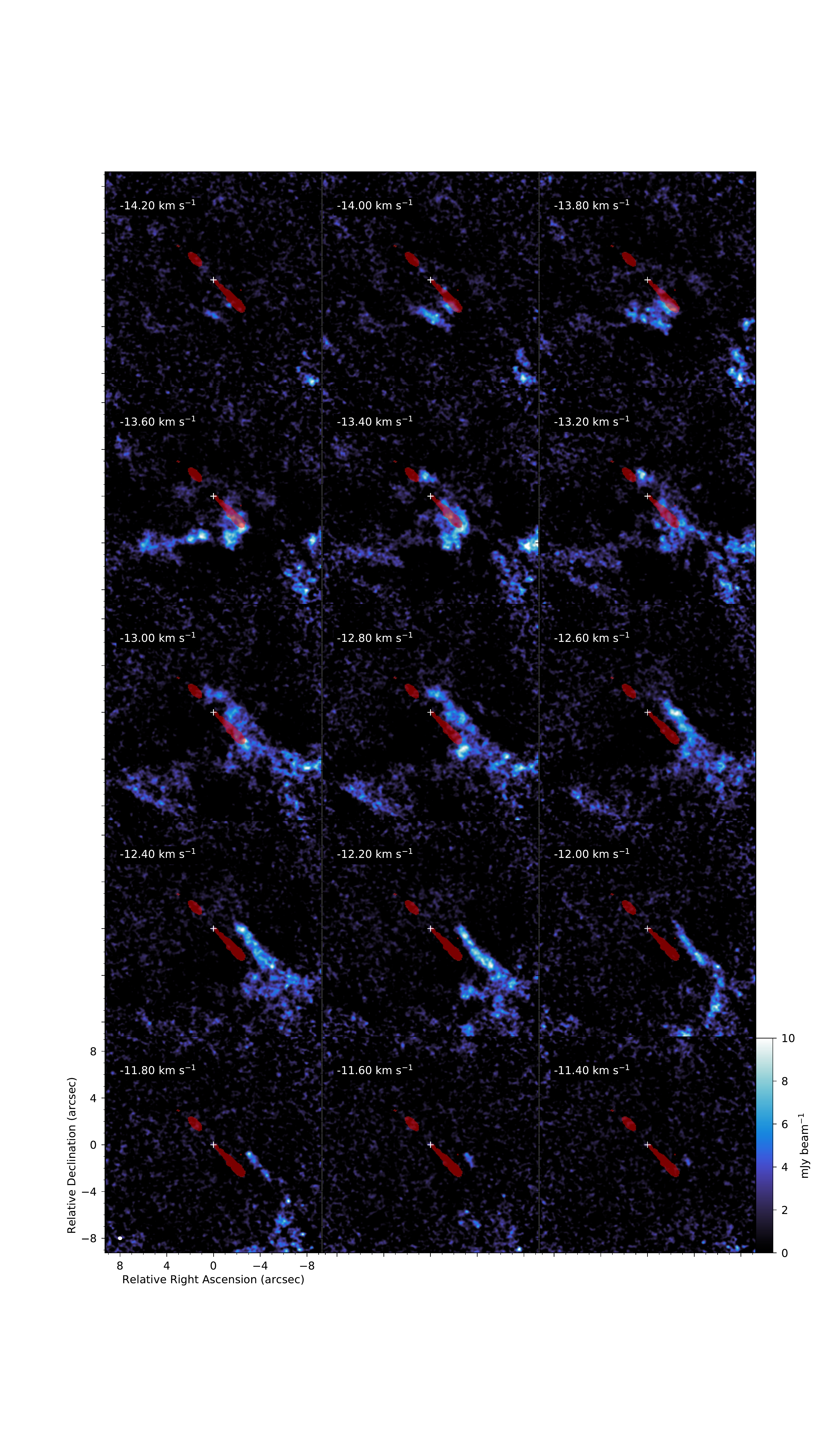}
\caption{$\mathrm{^{12}CO}$ J=2-1 channel maps with the jet traced in [SII] 673nm and H$\alpha$ from \citet{2020A&A...644A.149X} overlaid in red. Velocity axis is LSRK frame.}
\label{12co_chans3}
\end{figure*}

\begin{figure*}
    \centering
    \includegraphics[width=0.7\hsize]{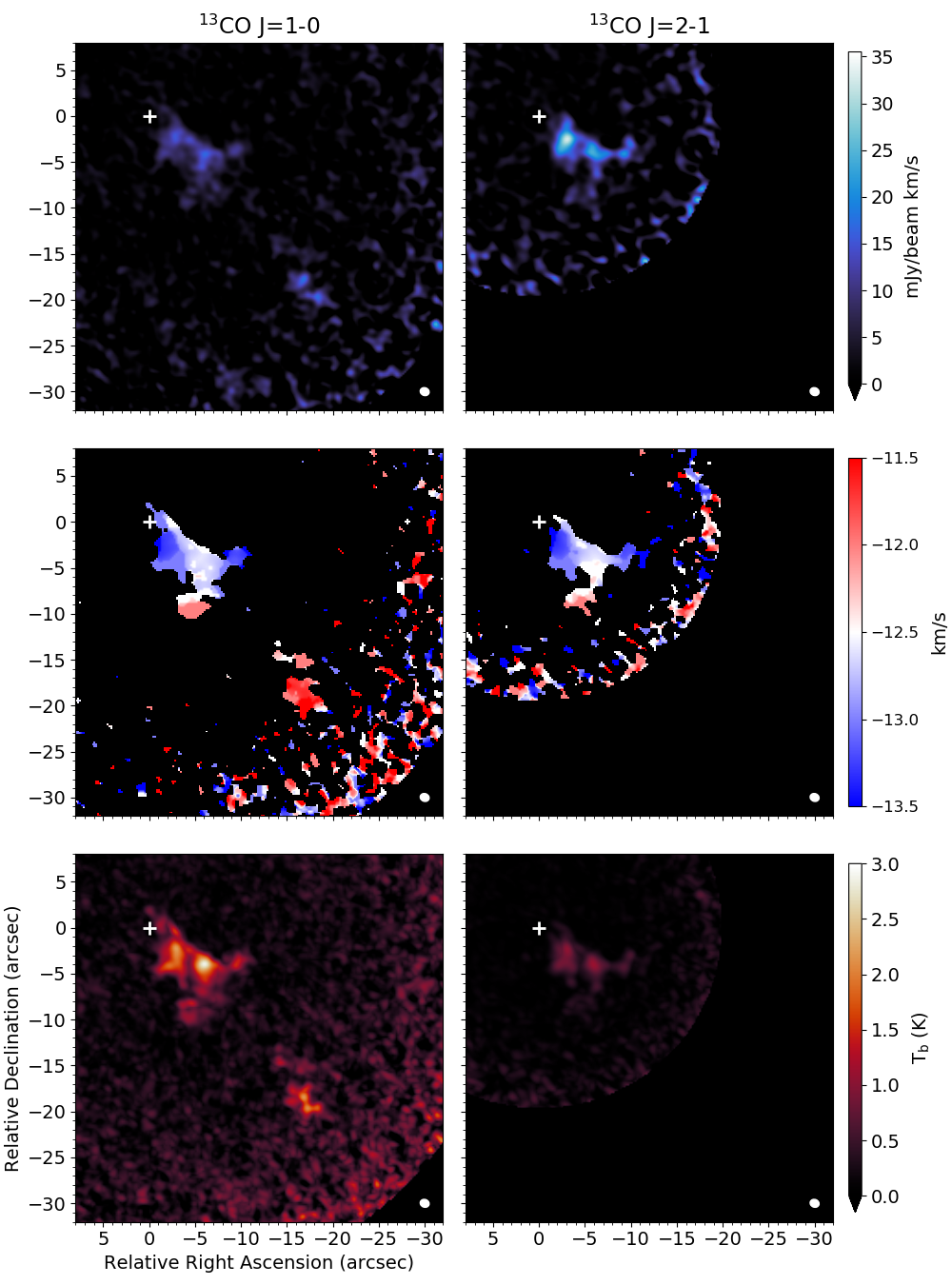}
    \caption{Integrated intensity (top), intensity-weighted velocity in LSRK frame. (middle) and peak brightness temperature (bottom) maps for $\mathrm{^{13}CO}$ $J=1-0$ and $J=2-1$ lines. 
    The beam size is shown by the ellipse in the bottom left corner of each image and star position is marked with a cross.}
    \label{13co_moments_app}
\end{figure*}

\section{12CO position-velocity diagrams}

\begin{figure*}
    \centering
    \includegraphics[width=0.7\hsize]{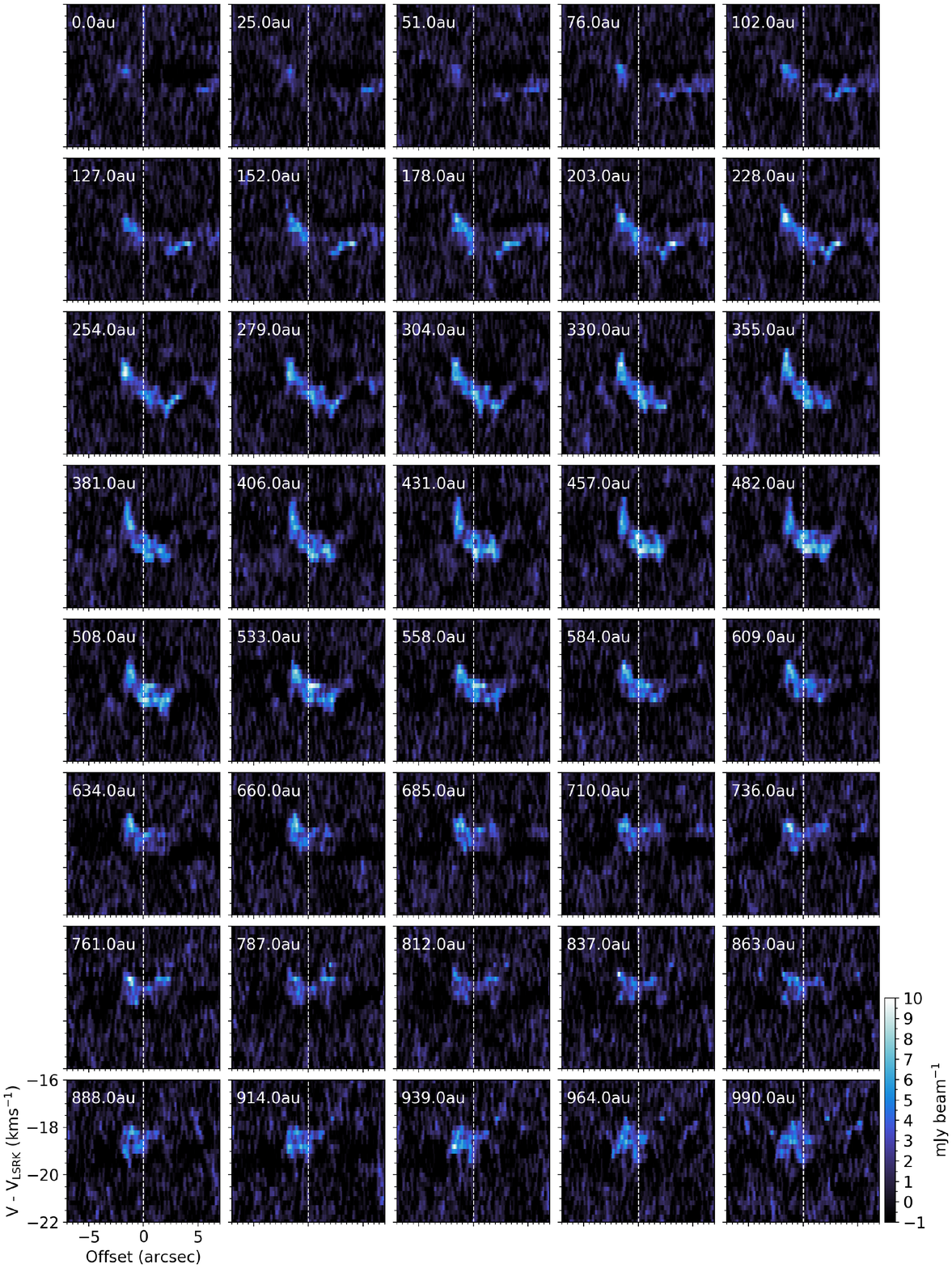}
   \caption{PV diagrams from the $\mathrm{^{12}CO}$ J=2-1 channel maps. Cuts are taken perpendicular to the jet axis and each PV diagram is averaged over a strip the width of the beam major axis with the center of the cut in au noted in the top left of each panel. The velocity axis has been corrected for the velocity of the source.}
    \label{fig:awesome_image}
\end{figure*}

\section{RADEX models with \citet{2013A&A...555A..73K} CO column density}

\begin{figure*}
    \centering
    \includegraphics[width=0.5\hsize]{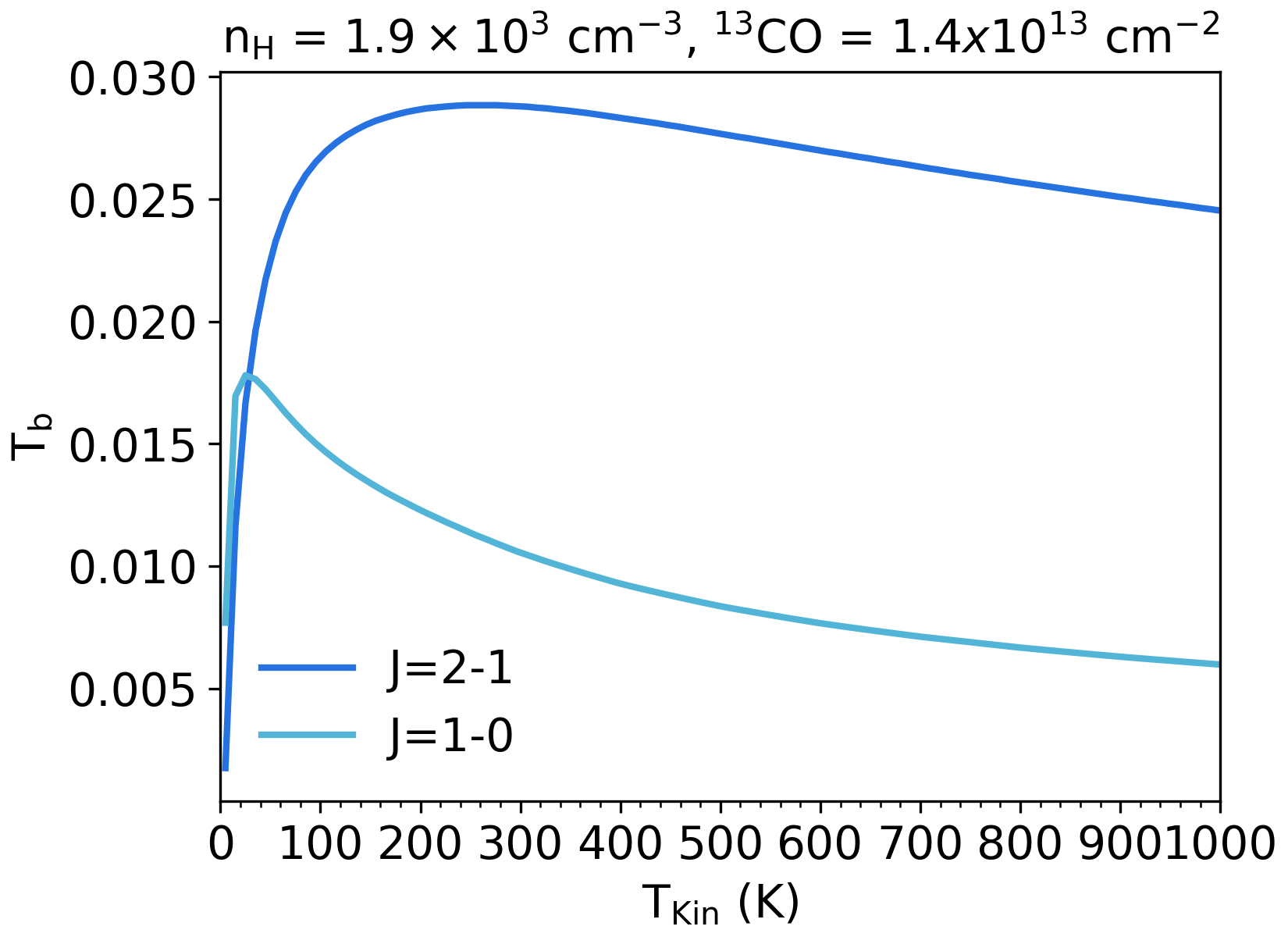}
    \caption{RADEX model brightness temperature ($\mathrm{T_b}$) with a $\mathrm{^{13}CO}$ column density of $10^{15}$/70 $\mathrm{cm^{-2}}$ to match values reported in \citet{2013A&A...555A..73K}.}
    \label{13co_tex_klaassen}
\end{figure*}

\bibliography{sample63}{}
\bibliographystyle{aasjournal}

\end{document}